\documentclass[aps,prl,reprint,nofootinbib,superscriptaddress,floatfix,longbibliography]{revtex4-2}
\usepackage[T1]{fontenc}
\usepackage{lmodern}
\usepackage{amsmath,amssymb,mathtools,bm,mathrsfs}
\usepackage{graphicx,booktabs,microtype,capt-of}
\usepackage[colorlinks=true,linkcolor=blue,citecolor=blue,urlcolor=blue]{hyperref}
\hypersetup{pdftitle={The Real EFThedron: A Moment Geometry for the Nonperturbative S-Matrix},pdfsubject={Main text, references, and one appendix},pdfkeywords={Real EFThedron; full unitarity; analyticity; moment problem; extremal measures; Hankel matrices}}
\graphicspath{{figures/}}
\newcommand{\avg}[1]{\left\langle #1\right\rangle}
\newcommand{\avgt}[1]{\left\langle #1\right\rangle_{\!\theta}}

\newcommand{\Ree}{\operatorname{Re}}
\newcommand{\Imm}{\operatorname{Im}}
\newcommand{\Disc}{\operatorname{Disc}}
\newcommand{\Tr}{\operatorname{tr}}

\newcommand{\be}{\begin{equation}}
\newcommand{\ee}{\end{equation}}
\newcommand{\bea}{\begin{aligned}}
\newcommand{\eea}{\end{aligned}}
\newcommand{\A}{\mathcal A}

\newcommand{\namedref}[2]{\hyperref[#2]{#1~\ref*{#2}}}

\usepackage{xcolor,xparse}
\newif\ifHRshowold
\HRshowoldtrue
\newcounter{HRoldblock}
\makeatletter
\NewDocumentCommand{\HRold}{+m}{%
  \ifHRshowold
    \begingroup
    \stepcounter{HRoldblock}%
    \edef\HRsavedequation{\number\value{equation}}%
    \color{red}%
    \hypersetup{linkcolor=red,citecolor=red,urlcolor=red}%
    \renewcommand{\theequation}{old}%
    \renewcommand{\theHequation}{HRold.\arabic{HRoldblock}.\arabic{equation}}%
    \let\label\@gobble
    \let\ltx@label\@gobble
    #1%
    \setcounter{equation}{\HRsavedequation}%
    \endgroup
    \space
  \fi
}

\makeatother

\begin{document}
\raggedbottom
\makeatletter
\immediate\write\@auxout{\string\citation{REControl}}
\makeatother
\title{The Real EFThedron: A Moment Geometry for the Nonperturbative S-Matrix}
\author{Laurentiu Rodina}
\affiliation{Beijing Institute of Mathematical Sciences
and Applications (BIMSA), Beijing 101408, China}

\author{Anna Tokareva}
\affiliation{School of Fundamental Physics and Mathematical
Sciences, Hangzhou Institute for Advanced Study, UCAS,
Hangzhou 310024, China}
\affiliation{Department of Physics, Blackett Laboratory, Imperial College London, SW7 2AZ London, UK}
\affiliation{International Centre for Theoretical Physics Asia-Pacific, Beijing/Hangzhou, China}
\author{Yongjun Xu}
\affiliation{School of Fundamental Physics and Mathematical
Sciences, Hangzhou Institute for Advanced Study, UCAS,
Hangzhou 310024, China}
\affiliation{Institute of Theoretical Physics, Chinese Academy of Sciences, Beijing 100190, China}
\affiliation{University of Chinese Academy of Sciences,
Beijing 100049, China}


\begin{abstract}
The S-matrix bootstrap restricts low-energy interactions through
unitarity, analyticity and crossing symmetry. For effective field theory
couplings, positivity bounds have a natural geometric formulation, the
EFThedron, in which the allowed couplings form the classical  positive moment problem. We show that this geometry extends to full nonlinear
unitarity. Multiplying the amplitude by fractional powers of the energy
turns its real part into a moment as well, so that unitarity becomes a
bounded complex moment problem. Analyticity becomes a set of linear
constraints requiring different kernels to reconstruct the same
couplings, analogous to the null constraints imposed by crossing
symmetry. 
In two dimensions the problem can be solved exactly, leading to
Castillejo--Dalitz--Dyson amplitudes as extremal measures of the moment problem. In four dimensions we obtain closed-form and
semi-analytic bounds for the massless EFThedron model and for
massive scalar scattering. For massive scalars,
the analyticity constraints improve on published dual
bounds,
reducing the gap to the primal region.  By
placing the real and imaginary parts of amplitudes on an equal footing,
the Real EFThedron provides a powerful framework combining unitarity,
analyticity and crossing, opening a new route to mapping the space of consistent quantum field
theories.
\end{abstract}
\maketitle

\section{Introduction}
\label{sec:introduction}
\label{sec:framework}
Low-energy interactions are constrained by the requirement that their
scattering amplitudes satisfy unitarity, analyticity and crossing
symmetry. The S-matrix bootstrap turns these requirements into bounds on
the couplings without specifying a microscopic model of the high-energy
dynamics~\cite{smatrixSnowmass,r18,r12,r5,r23,toolkit}.
The most widely studied bounds follow from dispersion relations.
Analyticity expresses a low-energy coefficient as an integral of the
imaginary part of the amplitude over high energies, and unitarity makes
the absorptive part positive~\cite{phamTruongPositivity,carvingOpenString,r1,scalarPositivity,r2,crossingCHD,crossingTWZ,multifieldPositivity}. Each
coupling is then a weighted average, or moment, of one positive but
otherwise unknown spectrum, and the allowed couplings solve a classical
moment problem. This defines the EFThedron~\cite{r7,r8}. Its geometry is
given by moments obeying positivity of Hankel matrices and linear relations due to crossing symmetry. The extremal theories living on the boundary have spectra made of a few delta functions, the
tree-level exchanges of individual heavy states.\footnote{The geometry of the EFThedron appears also in the  CFT bootstrap~\cite{cfthedron,modularGeometry}.} Related coefficient
bounds follow from geometric function theory~\cite{gftBieberbach,gftQFTEFT,gftPion}.

Positivity is only part of unitarity. For the partial wave of angular
momentum $\ell$, written as $S_\ell=1+if_\ell$, conservation of
probability requires $|S_\ell|\le1$, that is,
\begin{equation}
(\Ree f_\ell)^2+(\Imm f_\ell-1)^2\le1,
\label{eq:disk}
\end{equation}
with equality for purely elastic scattering. Keeping only
$0\le\Imm f_\ell\le2$ gives a moment problem with bounded density, the
deprojected EFThedron~\cite{r9,r15,gravityNonprojective}, but the disk contains more, since
it ties the real part to the imaginary part. The usual coefficient sum rules that express
couplings as moments involve only the imaginary part, so full unitarity
has been imposed in other ways: by reconstructing the real part through
Roy-type equations~\cite{r10,r4}, by parametrizing the amplitude in the
primal bootstrap~\cite{r18,r12,r19,r20,bootstrapFit,stringBootstrap,r4}, through numerical dual
functionals~\cite{r5,r6,hekruczenski,monolith,dualEFT,extremalHiggs,gumusThesis}, through dispersive iterations of unitarity and neural optimizers~\cite{tourkineZhiboedov,neuralBootstrap}, or by combining real-part sum
rules with full unitarity~\cite{r3}.

\begin{figure}[t]\centering\includegraphics[width=\columnwidth]{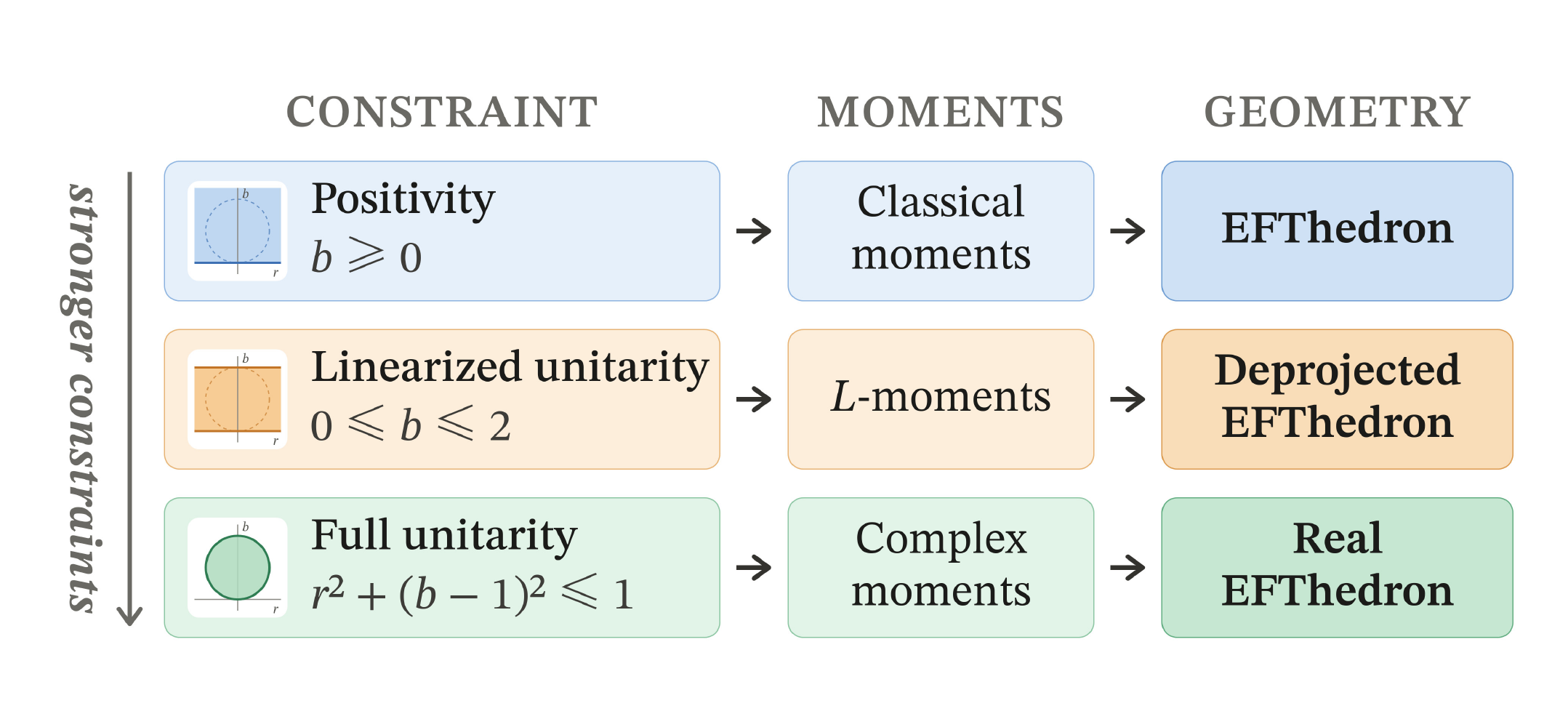}\caption{Three moment problems for a partial wave $f=r+ib$: positivity, linearized unitarity and full unitarity restrict $f$ to a half-plane, a strip and a disk. Their extremal spectra are delta functions, steps between $b=0$ and $b=2$, and pure phases on the unitarity circle.}\label{fig:hierarchy}\end{figure}

In this Letter we build full unitarity into the moment geometry.
Multiplying the amplitude by a fractional power $(1-x^2)^\alpha$ of the
energy variable before deforming the contour rotates the real part into
the discontinuity, so that both components of $f_\ell$ become moments
of one spectrum. Unitarity then defines a bounded complex moment problem
whose extremal spectra are pure phases, $|S_\ell|=1$, completing a
hierarchy in which positivity, linearized unitarity and full unitarity
have delta functions, steps and phases as extremal spectra
(Fig.~\ref{fig:hierarchy}). Analyticity becomes a set of linear null
constraints, obtained by demanding that different kernels reconstruct
the same couplings, in analogy with the null constraints of crossing
symmetry. The nulls cut linear slices through the complex moment
region, and we call the resulting geometry the \emph{Real EFThedron}.

These analytic nulls are closely related to the dispersion relation for
the real part of the amplitude and to the Roy equations~\cite{r10}. In
essence, they turn these dispersive constraints into linear, geometric
conditions on the moment space. Fixed-transfer averages also test
external energies beyond the compact window used in the dual
calculations of Refs.~\cite{r5,r26}. We combine these averages with
smeared Roy equations and harmonic nulls.

In two dimensions the approach is exact: $n$ analyticity constraints
determine the allowed region of the first $n+1$ couplings, and the
extremal measures are precisely Castillejo--Dalitz--Dyson (CDD)
amplitudes~\cite{r13,r11,r12}. 

In four dimensions we study the massless
EFThedron setting~\cite{r7,r9}, with light-particle loops neglected,
and massive scalars with the full physical threshold~\cite{r18,crossSection,r26}. In
both, the real part gives access to the constant coupling $g_0$, which
positivity cannot bound, and we derive closed-form inequalities as well
as stronger numerical bounds. For the massless model, bounds from full unitarity were obtained before with the primal bootstrap~\cite{r12} (see also \cite{extremalHiggs}). Here we provide dual bounds that follow from twice-subtracted dispersion relations, which are new for this model.

For massive scalars, the numerical bounds tighten the dual bounds of Refs.~\cite{r5,r26}, reducing their
separation from the primal region (Fig.~4). We proceed from analytic and semi-analytic bounds to numerical bounds and improve their constraining power in a systematic way, with unitarity imposed at all energies and spins. 


The present calculation uses selected families of these
constraints and does not exhaust their joint
optimization. Enlarging and combining the families
provides a systematic route to search for further
improvements.

\section{Unitarity as moments and analyticity as null constraints}
\label{sec:moment-matrix-bounds}

Dispersion relations express the low-energy coefficients of an
amplitude as integrals over its cuts. Writing the amplitude at
fixed momentum transfer in an energy variable $x$ for which the
direct- and crossed-channel cuts start at $x=\pm1$, and using
crossing to make it even, $\A(x)=\sum_n a_nx^{2n}$, Cauchy's
theorem gives
\begin{equation}
a_n=\frac2\pi\int_1^\infty\frac{\Imm\A(x+i0)}{x^{2n+1}}\,dx
\label{eq:dispersion}
\end{equation}
whenever the contributions from infinity can be dropped. Since
the amplitude takes complex-conjugate values on the two sides of
each cut, the real parts cancel in the discontinuity and only
$\Imm\A$ enters these sum rules.

To obtain coefficient sum rules involving the real part, we multiply
the amplitude by $K_\alpha(x)=(1-x^2)^\alpha$, with $\alpha$ real. Its own
discontinuity mixes the two components of the amplitude,
\begin{equation}
\begin{split}
\tfrac{1}{2i}\Disc(K_\alpha\A)
=(x^2-1)^\alpha\big[\cos \pi\alpha\,\Imm\A-\sin\pi\alpha\,\Ree\A\big],
\end{split}
\label{eq:fractional}
\end{equation}
reducing to the usual relation for $\alpha=0$ and isolating the
real part for $\alpha=1/2$. Kernels of this type have appeared in
continuous-moment sum rules~\cite{liuOkubo1967,olsson1968,continuousMoments}, which generalize finite-energy sum rules~\cite{dolenHornSchmid},
and in fractionally subtracted dispersion relations~\cite{r4}. In these works they were used to constrain the Regge behavior of the amplitude~\cite{continuousMoments} and to study the dependence of primal bounds on its asymptotic growth~\cite{r4}. Here we use them for a different purpose: they turn the real part of the amplitude into a moment of the same spectrum as the imaginary part, so that full unitarity can be imposed on the moments.

After a partial-wave expansion, every coefficient obtained in
this way becomes a linear functional of the complex partial
waves, so that the couplings take the form of complex moments
\begin{equation}
M_k=\sum_\ell\int\Imm(q_{k\ell}\,f_\ell)\,d\nu_\ell,\qquad |f_\ell-i|\le1,
\label{eq:complex-moments}
\end{equation}
with known complex weights $q_{k\ell}$ and positive reference measures
$\nu_\ell$. This is the analog of the classical moment problem
$m_k=\int z^kd\mu$ with $d\mu\ge0$, in which the positive density
has been replaced by a complex one confined to the unitarity disk.

\emph{Analyticity as linear constraints.---}
Because different kernels reconstruct the same coefficients, their results must agree, leading to linear constraints. We call these conditions \emph{analyticity null constraints}, by analogy with the crossing null
constraints that follow from crossing
symmetry~\cite{crossingCHD,crossingTWZ,csdrLocality,geometricCSDR}. They are not a new principle but a different way to implement the analyticity of the scattering amplitude.

A particularly useful family of such null constraints arises from comparing the ordinary
and square-root kernels. Introducing
$
z(x){=}\left(\frac{1-\sqrt{1-x^2}}{x}\right)^2$, with
$
\phi{=}2\arccos\frac1x
$
maps the two sides of the physical cut $x\ge1$ to $z_\pm=e^{\pm i\phi}$. Here $\phi$ labels energy, and
$\A_\phi=\A(\sec(\phi/2)+i0)$ is the upper-cut amplitude.
For integers $j\ge1$, Schwarz reflection combines the two cut
contributions into
\begin{equation}
\begin{split}
\mathcal H_j&\equiv
\frac2\pi\int_0^\pi d\phi\,\sin\phi\,
\Imm\!\big(e^{ij\phi}\A_\phi\big)\\
&=
\frac{1}{2\pi i}\oint dz\,
(z^{j-2}-z^j)\A(z),
\end{split}
\end{equation}
where $\Imm(e^{ij\phi} A)=\sin(j\phi)\Ree A+\cos(j\phi)\Imm A$. For $\A/x^2\to0$ and convergent endpoints, only $z=0$ contributes.
Since $\A(z)=a_0+\mathcal O(z)$, a residue occurs only for $j=1$,
giving
\begin{equation}
\mathcal H_j=a_0\,\delta_{j1}.
\label{eq:harmonic-identities}
\end{equation}

The first equation reconstructs the subtraction constant
$a_0=\A(0)$, while $j\ge2$ gives the \emph{harmonic nulls}. They are
closely related to the ordinary real-part dispersion relation at
physical energies $y>1$,
\begin{equation}
 \Ree\A(y+i0)=a_0+\frac{2y^2}{\pi}\operatorname{PV}
 \int_1^\infty\frac{\Imm\A(x+i0)}{x(x^2-y^2)}\,dx.
 \label{eq:real-pv}
\end{equation}
The principal value excludes the interval $(y-\epsilon,y+\epsilon)$ and takes
$\epsilon\to0$. The left-hand side is an unknown amplitude value,
and the kernel changes sign at $x=y$, so the equation does not
directly yield the usual positive-moment bounds.

The harmonics we introduced above can be also understood as weighted Fourier modes in energy of the difference
between the two sides. The complete sequence of harmonics recovers Eq.~\eqref{eq:real-pv}, while selected convergent sums
give energy averages of it. To compare, Roy equations instead average this
relation over scattering angle with partial-wave weights, and their
energy averages give smeared Roy equations~\cite{r10}. Appendix~\ref{app:roy-resummation}
gives the derivation and convergence conditions.

\emph{The Real EFThedron.---}
Unitarity defines a convex body of complex moments, which the
crossing and harmonic nulls cut by linear planes, defining the Real EFThedron. With finitely many
nulls this procedure gives an outer bound: more precisely, for a given set of nulls the Real EFThedron is the set of couplings for which there exist partial waves in the unitarity disk that reproduce these couplings and satisfy all the nulls. This set is convex, it shrinks when nulls are added, and it contains the physical region. In two dimensions it reproduces the exact region, while all four-dimensional bounds below are outer bounds. We compute bounds in two
complementary ways. 

\emph{A. Extremal measures.---}
In the first approach, the nulls are added with adjustable
multipliers to the combination of couplings to be bounded, and the
resulting integrand, $U\Ree f+V\Imm f$, is maximized at each
energy and spin. Its maximum, $V+\sqrt{U^2+V^2}$, lies on the
unitarity circle at the phase
\begin{equation}
\Ree f_*=\frac{U}{\sqrt{U^2+V^2}},\qquad
\Imm f_*=1+\frac{V}{\sqrt{U^2+V^2}},
\label{eq:phase}
\end{equation}
so that pure phases take over the role played by delta functions
in the positive moment problem and by steps in linearized
unitarity. Integrating this local maximum, summing over spins and
minimizing over the multipliers then yields the bound. This is the dual formulation of
the S-matrix bootstrap~\cite{lopez1975,bonnierLopezMennessier,lopezmennessier,r5,r6,hekruczenski,monolith,dualEFT,extremalHiggs}, used here with a new basis of functionals provided by the harmonic nulls. Geometrically, this construction is equivalent to the Minkowski sum over spins method described in \cite{r9}.

\emph{B. Moment matrices.---}
The second approach starts from the observation that full
unitarity is equivalent to positivity of the $2\times2$ matrix
\begin{equation}
W_\ell=\begin{pmatrix}\Imm f_\ell&\Ree f_\ell\\
\Ree f_\ell&2-\Imm f_\ell\end{pmatrix}\succeq0,
\qquad \Tr W_\ell=2.
\label{eq:W}
\end{equation}
Taking moments of $W_\ell$ in energy produces positive block
Hankel matrices. Their absorptive principal submatrices are the familiar EFThedron
Hankel matrices~\cite{r7,r8} of $\Imm f$ and $2-\Imm f$, now
linked by the real moments,
and the dispersion relations and nulls become linear equations
among their entries. Enlarging the matrices gives a systematic
hierarchy of constraints on the retained moment problem. Alternatively, Gram matrices with weights specially adapted to
the dispersion kernels can more efficiently lead to stronger bounds, including closed-form
bounds from small matrices~\cite{r15,r16,r21,r22}. Further details of both approaches are
given in Appendix~\ref{app:methods}.

\section{Two dimensions: exact finite matching}
\label{sec:two_dimensions}
In two dimensions there is no spin sum, and the coefficient region
can be determined exactly. For identical scalars of mass $m$, the
elastic S-matrix element and scattering amplitude are related by
\[
S(s)=1+\frac{iT(s)}{2\sqrt{s(s-4m^2)}},
\]
satisfying $SS^*\le1$ on the physical cut $s>4m^2$. We consider the pole-free, crossing-symmetric class in which $S$
is analytic on the
crossing-identified disk~\cite{r11,r12}. We define the couplings
at $s=2m^2$ by
\begin{equation}
T(s)=m^2\sum_{j=0}^\infty\frac{\Lambda_{2j}}{(2j)!}
\left(\frac{s-2m^2}{m^2}\right)^{2j}.
\label{eq:2d-normalization}
\end{equation}  Using the kernels of
Eq.~\eqref{eq:fractional} with $\alpha=0,-\tfrac12,-1$ we find
\begin{equation}
\begin{gathered}
-\tfrac14\Lambda_0=\avgt{\Imm f},\\
\Lambda_2+\tfrac14\Lambda_0=\avgt{\cos\theta\,\Imm f}
=\avgt{\sin\theta\,\Ree f},
\end{gathered}
\label{eq:2d-maps}
\end{equation}
where $f=T/[2\sqrt{s(s-4m^2)}]$ on the cut, and we defined the energy angle
$\cos\theta=1-8m^4/(s-2m^2)^2$ and the average
$\avgt F=\pi^{-1}\int_0^\pi F\,d\theta$. Since $\Lambda_2+\Lambda_0/4$ is reconstructed from both the imaginary
and the real part, analyticity requires the first analytic null constraint
\begin{equation}
N_1=\avgt{\sin\theta\,\Ree f-\cos\theta\,\Imm f}=0 .
\label{eq:N1}
\end{equation}

\begin{figure}[htbp]
\centering
\includegraphics[width=\columnwidth]{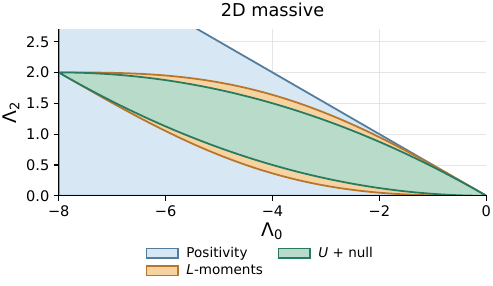}
\caption{Two-dimensional coupling regions from the imaginary-part
maps in Eq.~\eqref{eq:2d-maps}: positivity $\Imm f\ge0$ (blue),
bounded absorption $0\le\Imm f\le2$ (orange), and full unitarity
with the single null $N_1=0$ [Eq.~\eqref{eq:N1}] (green).
The green boundary is the exact lens in Eq.~\eqref{eq:lens}. The blue and orange boundaries are derived in Appendix~\ref{sec:end-2d}.}
\label{fig:2d}
\end{figure}

To find the boundary of the region allowed by full unitarity and
$N_1=0$, we bound $\Lambda_2+(1-\mu)\Lambda_0/4$ for each $\mu$ using
Eq.~\eqref{eq:phase}, with the null added through a multiplier
$\lambda$, so that $U=\lambda\sin\theta$ and
$V=\mu+(1-\lambda)\cos\theta$. For $|\mu|<1$ the optimal choice
$\lambda=(1-\mu^2)/2$ makes
\begin{equation}
U^2+V^2=\left[\frac{1+\mu^2}{2}+\mu\cos\theta\right]^2
\label{eq:square}
\end{equation}
a perfect square, so the extremal partial wave $f_*$ of
Eq.~\eqref{eq:phase} is a rational function of $\cos\theta$ and
$\sin\theta$. Remarkably, a single analytic null has selected an analytic intersection with the normally algebraic phase in Eq.~(\ref{eq:phase}). The corresponding S-matrix on the cut, $S_*=1+if_*$, has
$|S_*|=1$ and reads
\begin{equation}
S_*(\theta)=-\frac{\mu+e^{-i\theta}}{1+\mu e^{-i\theta}},
\label{eq:cdd-from-phase}
\end{equation}
which is the boundary value of a one-factor CDD amplitude~\cite{r13}. Plugging the extremal measure back into Eq.~(\ref{eq:2d-maps}), and eliminating $\mu$ yields the known exact
lens~\cite{r11,r12},
\begin{equation}
-8\le\Lambda_0\le0,\qquad
\frac{\Lambda_0^2}{32}\le\Lambda_2\le
-\frac{\Lambda_0}{2}-\frac{\Lambda_0^2}{32}.
\label{eq:lens}
\end{equation}

The same lens follows from two choices of Gram weights. Multiplying the two
rows and columns of $W$ in Eq.~\eqref{eq:W} by $\sqrt{1-\cos\theta}$
and $\sqrt{1+\cos\theta}$ and
averaging gives, by Eq.~\eqref{eq:2d-maps},
\begin{equation}
\begin{pmatrix}-\Lambda_0/2-\Lambda_2&\Lambda_2+\Lambda_0/4\\
\Lambda_2+\Lambda_0/4&2-\Lambda_2\end{pmatrix}\succeq0,
\label{eq:2d-principal-minors}
\end{equation}
whose determinant is the upper boundary in Eq.~\eqref{eq:lens};
exchanging the two weights gives the lower one.

At higher orders we obtain the nulls
\begin{equation}
N_j=\avgt{\sin(j\theta)\,\Ree f-\cos(j\theta)\,\Imm f}=0
\label{eq:finite-nulls}
\end{equation}
for $j=1,\ldots,n$. With unitarity, these $n$ nulls determine
exactly the region of the first $n+1$ couplings. Its boundary consists
of products of $n$ CDD factors
\begin{equation}
S_*(z)=e^{i\gamma}\prod_{j=1}^{q}\frac{z-z_j}{1-\bar z_jz},
\qquad z=-e^{-i\theta},
\label{eq:cdd-product}
\end{equation}
with $|z_j|<1$. Appendix~\ref{sec:end-2d} gives the
proof~\cite{schur17,schur18,r14}.

\section{Four dimensions: the massless EFThedron}
\label{eft4d:sec:massless_four_dimensions}
In both four-dimensional setups we use twice-subtracted dispersion
relations, assuming $\mathcal M(s,t)/s^2\to0$ as $|s|\to\infty$
at fixed $t$.

We first consider identical massless scalars in the standard
EFThedron setting~\cite{r7,r9}: the tree-level EFT amplitude holds
up to the cutoff, heavy states produce cuts starting at $s=1$, and
loops of the light particles are neglected. Following
Ref.~\cite{crossingCHD}, we define
\begin{equation}
\begin{gathered}
\mathcal M=16\pi\sum_{\ell\ {\rm even}}(2\ell+1)f_\ell(s)P_\ell(1+2t/s),\\
\mathcal M=g_0+g_2(s^2+t^2+u^2)+g_3stu+\cdots,
\end{gathered}
\label{eft4d:eq:eft-normalization}
\end{equation}
where $P_\ell$ are Legendre polynomials and each $f_\ell$ lies in the
unitarity disk. Unitarity is imposed only above the cutoff, along the heavy cut $s\ge1$. Below the cutoff only analyticity is imposed, which corresponds to the class of EFT amplitudes introduced in Ref.~\cite{r12}.

\emph{The coupling map.---}
At fixed $t$, the variable $z=(s+t/2)/(1+t/2)$ puts the cuts at
$|z|\ge1$ and makes the amplitude even, so the kernels of
Eq.~\eqref{eq:fractional} apply directly; for twice-subtracted
amplitudes they are admissible for $-1<\alpha\le n-1$ at order
$z^{2n}$. The ordinary relations give $g_2$ and $g_3$ in terms of
$\Imm f$, but $g_0$ is a subtraction constant that is not determined by
these absorptive sum rules, so positivity cannot constrain it. The
square-root kernel at order $z^4$ does contain it. With
$\avg{F}=\sum_{\ell\,{\rm even}}(2\ell+1)\int_1^\infty F_\ell\,ds/s^3$
and the energy angle $\cos\phi=2/s^2-1$, one finds at $t=0$
\begin{equation}
\avg{\Imm(e^{ij\phi}f)}=
\begin{cases}
g_2/16, & j=0,\\
g_0/128, & j=1,\\
0, & j\ge2.
\end{cases}
\label{eft4d:eq:main-eft-maps}
\end{equation}
 The higher modes give the harmonic
nulls $H_j\equiv\avg{\Imm(e^{ij\phi}f)}=0$. Standard crossing arguments add the first
crossing null $X_4\equiv\avg{L(L-8)\Imm f/s^2}=0$ with $L=\ell(\ell+1)$~\cite{crossingCHD}. Other
values of $\alpha$ and nonzero (non-forward) $t$ give further constraints. 

\begin{figure*}[t]
\centering
\includegraphics[width=1\textwidth]{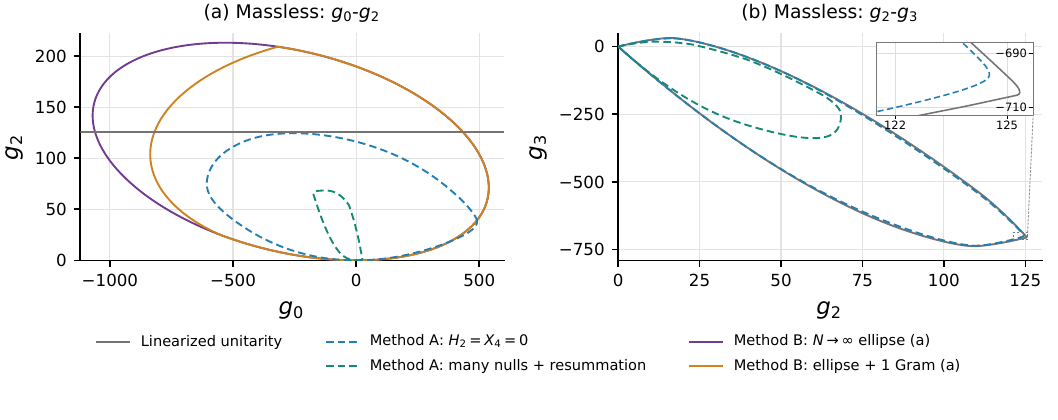}
 \caption{Massless coupling bounds.
Gray shows linearized unitarity with $X_4=0$.
In (a), solid purple shows the $N\to\infty$ Gram ellipse. Solid orange
intersects it with an additional Gram bound incorporating
$X_5=0$. Dashed blue uses the extremal measure method, and imposes full unitarity with $H_2=X_4=0$. Dashed green adds
nonforward and resummed relations.
The exact constraints and constructions are catalogued in
Table~\ref{app:tab:massless-current-curves}.}
\label{eft4d:fig:massless-comparison}
\end{figure*}

\emph{Method A: extremal measures.---}
The allowed region can be built one spin at a time. A single partial
wave $f_\ell(s)$ in the unitarity disk contributes a point to the space
of couplings and nulls, $(g_0,g_2,H_2,\ldots,X_4,\ldots)$, and as
$f_\ell$ varies these points fill a convex region $K_\ell$, whose
boundary is traced by the extremal phases of Eq.~\eqref{eq:phase}. 

 A full spectrum contributes one point from each spin, so it lies in the
Minkowski sum $\sum_\ell K_\ell$, and imposing the nulls cuts this sum
with the planes $H_2=X_4=\cdots=0$. The EFThedron has the same
structure, with each $K_\ell$ the cone spanned by the moment curve of
narrow spin-$\ell$ states, while in the deprojected
EFThedron~\cite{r9} $K_\ell$ is the image of $0\le\Imm f_\ell\le2$. In the Real EFThedron, each $K_\ell$ is the image of the unitarity disk. 

The simplest example is to bound $g_0$ using one harmonic null and one crossing null. For this we consider the space $(g_0,H_2,X_4)$, slice with $H_2=X_4=0$, obtaining a bound
$-606.4\lesssim g_0\lesssim491.1$.
Adding further analytic and crossing nulls, including nonforward
and resummed relations, gives the numerical bounds
\begin{equation}
 -173.32\lesssim g_0\lesssim24.85,
 \qquad 0\le g_2\lesssim68.387.
\end{equation}
The corresponding $(g_2,g_3)$ enclosure can be obtained in a similar way, improving on bounds
from linearized unitarity~\cite{r9}. The different regions
are shown in Fig.~\ref{eft4d:fig:massless-comparison}, and details are in Appendix~\ref{sec:app-massless-nonforward}.

\emph{Method B: moment matrices.---}
The EFThedron also has an algebraic description, in which moments of
the positive spectral density satisfy positivity of Hankel
matrices~\cite{r7,r8}. For a single spin, with $x=1/s$, define the
moments $I_k=\int_0^1x^{k+1}\Imm f_\ell\,dx$ and
$R_k=\int_0^1x^{k+1}\Ree f_\ell\,dx$. Positivity of $\Imm f_\ell$
gives the Hankel condition
\begin{equation}\begin{pmatrix}I_0&I_1\\I_1&I_2\end{pmatrix}\succeq0,
\end{equation}
linearized unitarity adds the same condition for $2-\Imm f_\ell$, and
full unitarity, in the matrix form of Eq.~\eqref{eq:W}, combines the
two into a single block Hankel matrix,
\begin{equation}
\begin{pmatrix}
I_0&R_0&I_1&R_1\\
R_0&1-I_0&R_1&\tfrac23-I_1\\
I_1&R_1&I_2&R_2\\
R_1&\tfrac23-I_1&R_2&\tfrac12-I_2
\end{pmatrix}\succeq0,
\label{eft4d:eq:block-hankel}
\end{equation}
obtained by integrating $(v_0+xv_1)^{T}W_\ell\,(v_0+xv_1)\,x\,dx$ for
arbitrary two-component vectors $v_0,v_1$. Rows and columns 1 and 3
form the EFThedron Hankel matrix, rows and columns 2 and 4 the one of
linearized unitarity, and the real moments in the remaining entries
couple the two. Once summed over spins, couplings and nulls are
linear in these moments, for example
$g_2=16\sum_\ell(2\ell+1)I_{0,\ell}$ and
$X_4=\sum_\ell(2\ell+1)L(L-8)I_{2,\ell}$, so matrix positivity
together with the nulls bounds the couplings, and larger matrices
give a systematic hierarchy. 

Weights adapted to the dispersion relations give Gram matrices
whose entries contain the couplings and nulls directly.
For example, using $X_4=H_2=\cdots=H_N=0$, one choice gives
\begin{equation}
\begin{pmatrix}g_2/16&h_N^{T}\\ h_N&2T_N\end{pmatrix}\succeq0,
\qquad
h_N=\Big(\frac{g_2}{16},\frac{g_0}{128},0,\ldots,0\Big)^{T},
\label{eft4d:eq:gram-matrix}
\end{equation}
where $T_N$ is a known reference matrix. Positivity gives an
ellipse in the $(g_0,g_2)$ plane. For the weights used here,
its $N\to\infty$ limit can be computed analytically:
\begin{equation}
g_0^2+4.9868\,g_0g_2+57.0222\,g_2^2
\lesssim10825.064\,g_2.
\label{eft4d:eq:ellipse-limit}
\end{equation}
This gives a boundary described by a single quadratic inequality. A further $4\times4$ Gram construction uses
$H_2=X_4=X_5=0$
\begin{equation}
 \begin{pmatrix}
 g_2/16 & g_2/16 & g_0/128 & 0\\
 g_2/16 & 2\mu_0 & 2\mu_1 & 2\mu_2\\
 g_0/128 & 2\mu_1 & 2\mu_0 & 2\mu_1\\
 0 & 2\mu_2 & 2\mu_1 & 2\mu_0
 \end{pmatrix}\succeq0,
 \label{eft4d:eq:additional-gram}
\end{equation}
where $\mu_0\simeq7.09817$, $\mu_1\simeq0.43304$,
and $\mu_2\simeq-0.93491$ are fixed reference constants.
Its determinant gives another ellipse. Intersecting it with
the $N\to\infty$ ellipse gives the orange boundary in
Fig.~\ref{eft4d:fig:massless-comparison}(a).
The weights and exact constants are given in Appendices~\ref{app:massless-matrices} and~\ref{app:massless-x5-grams}.

\section{Four dimensions: physical massive scattering}
\label{eft4d:sec:massive_four_dimensions}

The same geometry applies to massive scattering, now with the
physical threshold at $s=4m^2$. We consider identical scalars with crossing symmetry, no
subthreshold poles,  retain the
same twice-subtracted assumption, and set $m=1$. We use the established fixed-transfer analyticity domain~\cite{jinmartin,martin65,martin66}: at fixed transfer in the range $-4<t<4$ the amplitude is analytic in the complex $s$ plane apart from the direct and crossed cuts and satisfies a twice-subtracted dispersion relation, as in Ref.~\cite{r4}. We use in addition the endpoint and convergence conditions stated in Appendix~\ref{app:fractional-kernels}. The couplings are defined at the
crossing-symmetric point $s=t=u=4/3$,
\begin{equation}
 \frac{\mathcal M}{32\pi}
 =g_0^{\rm m}
 +g_2^{\rm m}(\bar s^2+\bar t^2+\bar u^2)+\cdots,
 \label{eft4d:eq:massive-coefficients}
\end{equation}
with $\bar s=s-4/3$ and similarly for $\bar t,\bar u$. The couplings $g_0^m$ and $g_2^m$ correspond to $c_{0,0}$ and
$c_{1,0}$ of Ref.~\cite{r26}. The partial waves are
\begin{equation}
 \mathcal M=\frac{16\pi}{\rho}
 \sum_{\ell\ {\rm even}}(2\ell+1)f_\ell
 P_\ell\!\left(1+\frac{2t}{s-4}\right),
\label{eft4d:eq:main-massive-measure}
\end{equation}
with $\rho=\sqrt{1-4/s}$ and each $f_\ell$ in the unitarity disk. 

\emph{The coupling map.---}
The symmetric point is reached most directly at fixed $t=4/3$. The
variable $z=3\bar s/8$ places the cuts at $|z|\ge1$ and the
symmetric point at $z=0$. The ordinary and square-root
kernels then give the same Fourier structure as in the massless case,
\begin{equation}
\avg{p_\ell\,\Imm(e^{ij\phi_{\rm m}}f)}_{\rm m}=
\begin{cases}
g_2^{\rm m}, & j=0,\\
9g_0^{\rm m}/512, & j=1,\\
0, & j\ge2,
\end{cases}
\label{eft4d:eq:main-massive-maps}
\end{equation}
where $p_\ell=P_\ell(\bar s/(s-4))/\bar s^3$, $\cos\phi_{\rm m}{=}128/(9\bar s^2){-}1$ is the energy angle, and we defined the spectral average $\langle F\rangle_{\rm m}{=}\sum_{\ell\ {\rm even}}(2\ell{+}1)
 \int_4^\infty\frac{F_\ell(s)}{2\pi\rho}\,ds$. 

The symmetric-point harmonic nulls are
$\mathcal H_j^*{=}\avg{p_\ell\Imm(e^{ij\phi_{\rm m}}f)}_{\rm m}{=}0$
for $j\ge2$. It is also useful to define the forward moments
\begin{equation}
\widetilde H_j(0)=
\left\langle\frac{32}{(s-2)^3}
\Imm(e^{ij\vartheta_{\rm m}}f)\right\rangle_{\rm m}
\end{equation}
where $\cos\vartheta_{\rm m}=\frac8{(s-2)^2}-1$. These harmonics vanish for $j\ge2$.  The first crossing null at the symmetric point is
$\widetilde X_4$.

\begin{figure}
    \centering
    \includegraphics[width=1\linewidth]{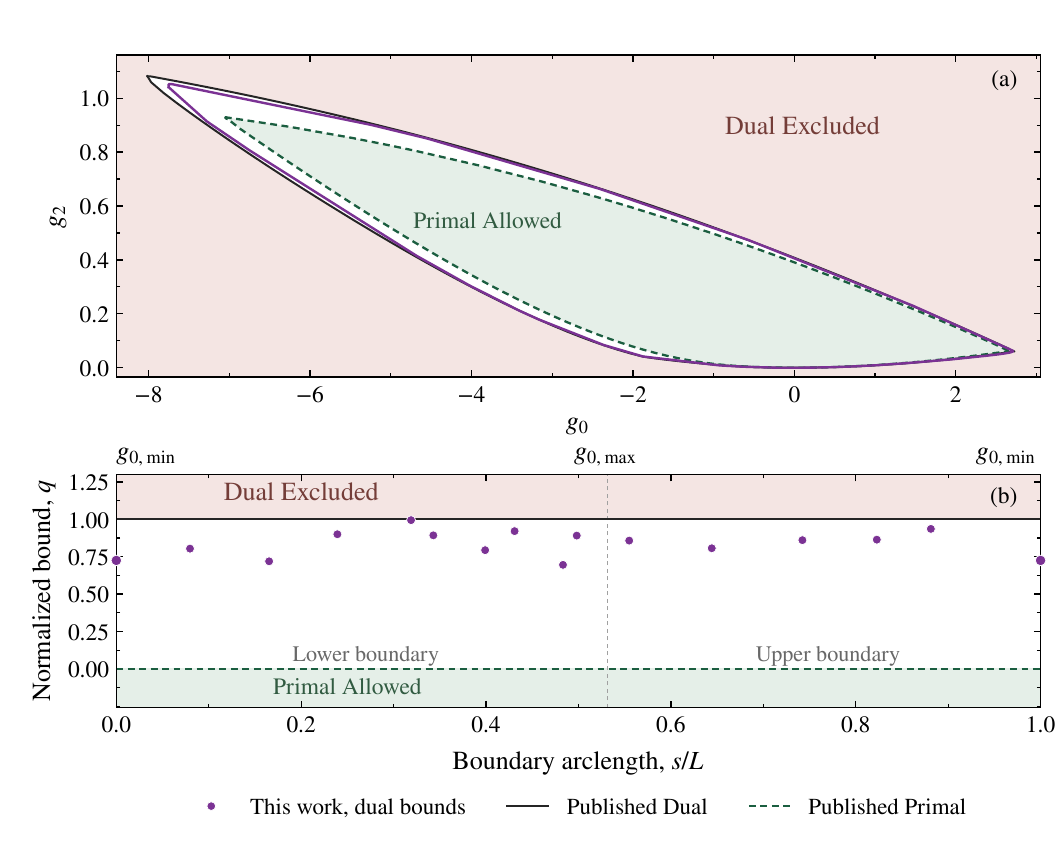}
    \caption{Massive scalar bounds in four dimensions.
(a) Our certified enclosure is shown in purple. The published dual
(black) and primal (green dashed) curves are from Ref.~\cite{r26}.
(b) Certified inequalities compared with those curves:
$q=1-\delta_C/\delta_P$, with $\delta_C,\delta_P$ the signed inward
normal displacements from the dual curve in axis-range-normalized
coordinates. The dual and primal have $q=1$ and $q=0$. Arclength runs counterclockwise from minimum $g_0^{\rm m}$.}
    \label{eft4d:fig:massive-comparison}
\end{figure}

\emph{Method A: extremal measures.---}
The construction is identical to the massless case, now in the space
$(g_0^{\rm m},g_2^{\rm m},\widetilde H_2(0),\widetilde X_4,\ldots)$.
With only $\widetilde H_2(0)$ and $\widetilde X_4$, we obtain $g_0^{\rm m}\lesssim4.34$, with $g_2^{\rm m}$
unbounded. 

By including more constraints, in particular non-forward and resummed harmonics, we obtain significantly stronger, yet still rigorous, bounds
\begin{equation}
\begin{gathered}
-7.751502\le g_0^{\rm m}\le2.705955,\\
0\le g_2^{\rm m}\le1.052351.
\end{gathered}
\label{eft4d:eq:certified-massive-bounds}
\end{equation}
The resulting
exclusions improve on the published dual boundaries of Ref.~\cite{r26}, as shown in Fig.~\ref{eft4d:fig:massive-comparison}. Appendix~\ref{app:massive-resummed-registry} lists
all retained constraints, 
and explains how the bounds are rigorously certified.


\emph{Method B: moment matrices.---}
Mass only changes the reference measure, so the same Gram construction
applies. For a simple closed family, use the forward harmonic moments
$\widetilde H_2(0)=\cdots=\widetilde H_N(0)=0$. The adapted weights give
\begin{equation}
 \begin{pmatrix}g_2^{\rm m}&\widetilde H_1(0)\mathbf1^T\\
 \widetilde H_1(0)\mathbf1&2T_N\end{pmatrix}\succeq0,
 \label{eft4d:eq:massive-gram}
\end{equation}
an $N\times N$ matrix with known reference block $T_N$ and
$\mathbf1$ of length $N-1$. Here $\widetilde H_1(0)$ is the forward moment;
its difference from $g_0^{\rm m}$ is a bounded absorptive contribution.
Eliminating it gives
\begin{equation}
 \begin{gathered}
 -\frac{128}{27}g_2^{\rm m}-\sqrt{2K_Ng_2^{\rm m}}
 \le 
 g_0^{\rm m}\le-\frac{64}{27}g_2^{\rm m}+\sqrt{2K_Ng_2^{\rm m}},
 \end{gathered}
 \label{eft4d:eq:massive-gram-bound}
\end{equation}
where $ K_N=(\mathbf1^TT_N^{-1}\mathbf1)^{-1}$. Increasing the matrix size strengthens both bounds, with the analytic
limit $K_\infty\approx233.75$. 
The derivation is given in Appendix~\ref{app:massive-forward-matrices}. We leave a systematic study of Gram matrices for the massive scalars for future work.

\section{Outlook}
We have formulated the nonperturbative S-matrix bootstrap as the geometry of a moment problem. Fractional dispersive kernels make the real part of the amplitude a moment of the same spectrum as its imaginary part, so that full unitarity becomes a bounded complex moment problem. This extends the EFThedron and completes the progression from positivity through linearized to full unitarity, in which the extremal spectra evolve from delta functions to steps to pure phases. The crucial ingredients are the analyticity nulls and the related harmonics, which resum to the real-part dispersion relation, contain the Roy equations, and constrain the real part at all energies. 

In two dimensions finitely many such nulls give the exact coupling regions. In four dimensions they lead to closed-form bounds from Gram matrices and to stronger numerical bounds from extremal measures. For massive scalars, the certified inequalities improve on portions of the published dual boundaries~\cite{r5,r26}, narrowing the gap to the primal results~\cite{r18,r26} while relying only on the proven analyticity requirements. The constraint sets used here are far from exhaustive, so a systematic optimization, including crossing-symmetric dispersion relations~\cite{aubersonKhuri,csdrSinhaZahed,geometricCSDR}, should strengthen these bounds further.

More broadly, placing real and imaginary parts on an equal footing opens the application of moment geometry to theories with light particles ~\cite{IRpositivity,runningEFT}, and, once forward and infrared limits are controlled, to gauge and gravitational scattering~\cite{gravitationalPositivity,sharpSwampland,einsteinCausality,finiteMPL,samplingGraviton}. Fractional kernels can be easily adjusted to the arbitrary choice of Regge bound assumptions, making them a promising tool for dual bootstrap of the gravitational EFTs.

\vspace{0.5cm}
\paragraph{Acknowledgments} Authors are grateful to Joan Elias-Miro, Denis Karateev, Guangzhuo Peng, Piotr Tourkine for valuable discussions. LR is supported by the National Natural Science Foundation of China General Program No. 12475070 and the Beijing Natural Science Foundation International Scientist Project No. IS24014. The work of AT was supported by the National Natural Science Foundation of China (NSFC) under grants No. 12547104, No. 12505091, and Zhejiang Provincial Natural Science Foundation of China under Grant Z-YC-QN-s17p-000738. AI tools were used to assist with numerical methods, optimization and consistency checks, and manuscript editing.

\paragraph*{Data availability.}
The numerical code, including constraint kernels and verification
scripts, is available at
\url{https://github.com/yongjunx23-del/The-Real-EFThedron}.
The numerical data underlying the figures are provided as ancillary
files. 

\nocite{r20,r21,r24,r25,toolkit,r26}
\bibliography{references}

\clearpage
\onecolumngrid
\appendix
\setcounter{section}{0}
\setcounter{secnumdepth}{2}
\renewcommand{\thesection}{A\arabic{section}}
\renewcommand{\thesubsection}{\thesection.\arabic{subsection}}
\makeatletter\renewcommand{\p@subsection}{}\makeatother
\renewcommand{\theHsection}{appendix.\arabic{section}}
\renewcommand{\theHsubsection}{\theHsection.\arabic{subsection}}
\numberwithin{equation}{section}
\renewcommand{\theequation}{\thesection.\arabic{equation}}
\renewcommand{\theHequation}{\theHsection.\arabic{equation}}
\setcounter{figure}{0}
\setcounter{table}{0}
\renewcommand{\thefigure}{A\arabic{figure}}
\renewcommand{\thetable}{A\arabic{table}}
\renewcommand{\theHfigure}{appendix.\arabic{figure}}
\renewcommand{\theHtable}{appendix.\arabic{table}}

\section{Moment matrices, extremal measures, and null constraints}
\label{app:methods}
\label{app:matrix-construction}
\label{app:extremal-measures}

Hereafter we use the notations $f_\ell=r_\ell+ib_\ell$. In what follows, the reference measures and weights
are known functions, while the same unknown spectrum enters every dispersion relation.
All integrals below, including their spin sums, are required to converge.

\subsection{Hankel and Gram matrices}
\label{app:hankel-gram}

We work at a fixed spin and suppress its index throughout.
We choose $x\in[0,1]$ and a finite positive reference
measure $d\nu$. We define
$I_k=\int x^k b\,d\nu$, $R_k=\int x^k r\,d\nu$, and
$\nu_k=\int x^k\,d\nu$. Positivity of the partial-wave
imaginary part $b$ gives, for every $a,c$,
\begin{equation}
 \int(a+cx)^2b\,d\nu\ge0
 \quad\Longleftrightarrow\quad
 \begin{pmatrix}I_0&I_1\\I_1&I_2\end{pmatrix}
 \succeq0,
 \qquad I_1^2\le I_0I_2.
 \label{eq:moment-hankel-example}
\end{equation}
This is a Hankel matrix, whose entries depend
only on the sum of the row and column indices,
corresponding to monomial test functions.
Applying the same test to $2-b$ implements the
boundedness of the absorption part of the amplitude.
Full unitarity combines these tests through the matrix
$W$ of Eq.~\eqref{eq:W}. Its moments and the block
Hankel hierarchy are
\begin{equation}
 Y_k=\int x^kW\,d\nu=
 \begin{pmatrix}I_k&R_k\\R_k&2\nu_k-I_k\end{pmatrix},
 \qquad
 \mathsf H_d=[Y_{i+j}]_{i,j=0}^{d}\succeq0,
 \qquad
 \mathsf L_{d-1}
 =[Y_{i+j+1}-Y_{i+j+2}]_{i,j=0}^{d-1}
 \succeq0.
 \label{app:eq:matrix-moments}
\end{equation}
The first condition is an integral of $v(x)^T W v(x)$
for a vector polynomial $v(x)$ of degree
at most $d$.
The second, for $d\ge1$, corresponds to an insertion
of $x(1-x)$. For example, the condition
$Y_0\succeq0$ gives
\begin{equation}
 R_0^2\le I_0(2\nu_0-I_0),
 \qquad
 R_0^2+(I_0-\nu_0)^2\le\nu_0^2.
 \label{app:eq:averaged-disk}
\end{equation}
Thus, averaging preserves the full unitarity condition
written as a semidefinite matrix. At $d=1$ the matrix
is $4\times4$ and additionally relates different
energy moments, as in
Eq.~\eqref{eft4d:eq:block-hankel}.
When the dispersion kernels are expressed in the
chosen moment basis, low-energy couplings are linear
combinations of moments across different spins.
Dispersion relations and nulls supply more linear
equations. Eliminating unfixed moments gives
inequalities on couplings.

More generally, Hankel matrices are special cases
of Gram matrices, obtained by choosing monomial test
functions.
For general real two-component weights $\psi_a$,
the same argument gives
\begin{equation}
  G_{ab}=\int\psi_a^TW\psi_b\,d\nu,
 \qquad G\succeq0.
 \label{app:eq:weighted-gram}
\end{equation}
This is a Gram matrix.
Indeed, for arbitrary real coefficients $c_a$,
its quadratic form
$\sum_{a,b}c_aG_{ab}c_b
=\int(\sum_a c_a\psi_a)^T
W(\sum_b c_b\psi_b)\,d\nu$
is nonnegative.
Monomial weights $x^k(1,0)^T$ and $x^k(0,1)^T$
reproduce the block Hankel matrices.
More general weights can be chosen to match
the kernels appearing in particular dispersion relations.
For example, the weights
$\psi_1=(p,0)^T$ and $\psi_2=(0,q)^T$
with real functions $p$ and $q$ give
\begin{equation}
 \begin{pmatrix}
 \int p^2b\,d\nu&\int pqr\,d\nu\\
 \int pqr\,d\nu&\int q^2(2-b)\,d\nu
 \end{pmatrix}\succeq0.
 \label{eq:moment-gram-example}
\end{equation}
The determinant bounds the real-part moment by the
two diagonal moments. The 2D weights
$p=\sqrt{1-\cos\theta}$,
$q=\sqrt{1+\cos\theta}$
are chosen such that
$pq=\sin\theta$ for $\theta\in[0,\pi]$,
turning the mixed entry into the harmonic moment in
$N_1$, giving the lens derived in
Appendix~\ref{sec:end-2d}.

The Hankel hierarchy thus provides a systematic
polynomial basis, while general Gram weights allow the
positivity constraints to be adapted to particular
dispersion kernels. In the dual formulation, a candidate
coupling bound can lead to a polynomial matrix condition
$M(x;y)\succeq0$ for all $x\in[0,1]$, where $y$
denotes the dual variables. For a symmetric $2\times2$
polynomial matrix of degree at most $2d$, this condition
admits the exact sum-of-squares representation
\begin{equation}
 \begin{aligned}
 M(x;y)
 ={}&V_d(x)^TQ_0V_d(x)+x(1-x)V_{d-1}(x)^TQ_1V_{d-1}(x),
 \end{aligned}
 \qquad Q_0,Q_1\succeq0,
 \label{app:eq:pmp-sdp}
\end{equation}
with
$V_d(x)=(1,x,\ldots,x^d)^T\otimes\mathbb I_2$.
Matching polynomial coefficients gives linear equations
in $y$ and the entries of the positive-semidefinite
Gram matrices $Q_0$ and $Q_1$.
The polynomial matrix program (PMP) is thereby converted
into an SDP. This provides a dual positivity certificate
complementary to the moment constraints above.

In general, completeness of the moment hierarchy for a
fixed measure does not by itself prove convergence of a
truncated spin sum or a finite null set.
The choice of dispersion relations, test functions,
and null constraints can also affect the strength of
bounds at finite truncation.
The convergence can be checked by comparing the bounds
to explicitly constructed physical primal
amplitudes~\cite{aubersonEpeleSimao,aubersonMennessier,r18,r12,r26}.

\subsection{Extremal measures and Minkowski sums}
\label{app:extremal-slice}
Let $L[f]$ be a linear combination of couplings that we want to bound.
Let $N_a[f]=0$ be the null constraints that we impose, where the index
$a$ labels them. Both $L[f]$ and $N_a[f]$ are linear functionals of
the partial waves. For arbitrary real multipliers $\lambda_a$ we write
$L+\sum_a\lambda_aN_a=\sum_\ell\int(U_\ell r_\ell+V_\ell b_\ell)d\nu_\ell$.
This relation defines the real kernels $U_\ell$ and $V_\ell$. They are
both known functions of the energy within the chosen set of dispersion relations and constraints, and they depend linearly on the Lagrange
multipliers. At each energy and spin we maximize the integrand over
the unitarity disk, which means that the extremal partial waves have a form of pure phases. This gives an equation 
\begin{equation}
 L[f]\le B(\lambda):=
 \sum_\ell\int\!\left(V_\ell+\sqrt{U_\ell^2+V_\ell^2}\right)d\nu_\ell .
 \label{app:eq:disk-support}
\end{equation}
The maximum is reached at the extremal phase of Eq.~\eqref{eq:phase}.
Therefore, every choice of the multipliers gives a valid inequality.
We minimize $B(\lambda)$ over the multipliers, in order to obtain the strongest
inequality. At the minimum the derivative of $B(\lambda)$ with respect
to $\lambda_a$ vanishes. This derivative is the null $N_a$ evaluated
on the extremal phase. So, the extremal phase satisfies the
null constraints, provided that $B(\lambda)$ is a smooth function at the minimum.

This construction has a geometric meaning. The support function of a
convex set in a given direction is the best upper bound on the chosen combination of couplings. Each spin contributes
a convex region $K_\ell$, as described in the main text, and the
integral in Eq.~\eqref{app:eq:disk-support} at fixed $\ell$ is the
support function of this region. The support function of a Minkowski
sum is the sum of the support functions of its terms.
Equation~\eqref{app:eq:disk-support} is therefore the dual description
of the following construction. We build the regions $K_\ell$, we take
their Minkowski sum, and we intersect the sum with the planes defined
by the nulls. The nulls are imposed on the sum over spins and not on
each spin separately, because the contributions of different spins can
cancel in a null.

A finite set of inequalities of the form~\eqref{app:eq:disk-support}
defines a convex region that contains the physical region. In what follows we also call
this region an outer enclosure. Its boundary need not be reached by
an analytic amplitude. For example, the extremal phases are purely
elastic at all energies, whereas in four dimensions scattering implies
particle production~\cite{aks1965}. In the problem with all spins,
the contribution of the spins that are not integrated explicitly must
be bounded analytically. A useful estimate for $V_\ell<0$ is
$V_\ell+\sqrt{U_\ell^2+V_\ell^2}\le U_\ell^2/(2|V_\ell|)$, which
leads to a convergent sum over spins in our applications. The bounds
can be improved in two independent ways. We can enlarge the set of
null constraints, and we can strengthen the way in which unitarity is imposed.

\subsection{Fractional kernels and coefficient elimination}
\label{app:fractional-kernels}
In this subsection and in Appendix~\ref{app:roy-resummation} the
symbol $x$ denotes the energy variable of Eq.~\eqref{eq:dispersion}.
At fixed momentum transfer $t$ the amplitude is an even function
$\mathcal A_t(x)=\sum_n a_n(t)x^{2n}$ with cuts at $|x|\ge1$. For both the
massless and massive models this variable is called $z$ in the main
text. We multiply the amplitude by the fractional kernel
$(1-x^2)^\alpha$ of Eq.~\eqref{eq:fractional}, and we extract the
coefficient of $x^{2n}$ by a contour integral around the origin. We
denote the result by $\mathcal C_n(\alpha,t)$. The expansion of the
kernel in powers of $x$ gives the first line of the following equation.
The deformation of the contour onto the cuts gives the second line,
in which the amplitude is evaluated on the upper side of the right
cut,
\begin{equation}
 \begin{aligned}
 \mathcal C_n(\alpha,t)
 &=\frac1{2\pi i}\oint\frac{(1-x^2)^\alpha\mathcal A_t(x)}{x^{2n+1}}\,dx
 =\sum_{q=0}^{n}(-1)^q\binom\alpha q a_{n-q}(t)\\
 &=\frac2\pi\int_1^\infty\frac{(x^2-1)^\alpha}{x^{2n+1}}
 [\cos(\pi\alpha)\Imm\mathcal A_t-\sin(\pi\alpha)\Ree\mathcal A_t]\,dx.
 \end{aligned}
 \label{eq:end-alpha-coefficients}
\end{equation}
\begin{figure}[htb]
\centering
\includegraphics[width=0.45\textwidth]{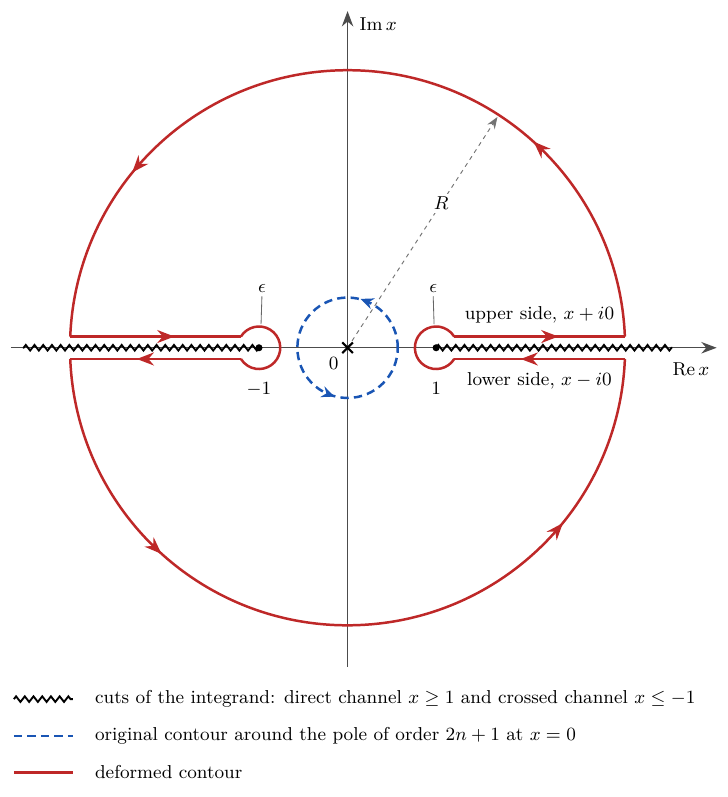}
\caption{The figure shows the contour of integration of
Eq.~\eqref{eq:end-alpha-coefficients} in the complex plane of the
energy variable $x$. The dashed circle is the original contour around
the pole at $x=0$, and the solid line is the deformed contour, which
runs along both sides of the two cuts. The arcs of radius
$R\to\infty$ and the circles of radius $\epsilon\to0$ around the
points $x=\pm1$ give no contribution under the conditions stated in
the text. On the upper and lower sides of the right cut,
$x\pm i0$, the kernel is equal to
$(1-x^2)^\alpha=(x^2-1)^\alpha e^{\mp i\pi\alpha}$. The left cut,
$x\le-1$, gives the same contribution as the right cut, because
$\mathcal A_t(-x)=\mathcal A_t(x)$.}
\label{app:fig:contour}
\end{figure}
The kernel $(1-x^2)^\alpha$ is real and positive for $-1<x<1$. We
assume that $\mathcal A_t/x^2\to0$ at infinity and that
$\mathcal A_t=O((x-1)^{-\gamma})$ at threshold with some exponent
$\gamma$. The deformation of the contour is then allowed for
$\gamma-1<\alpha\le n-1$. The lower limit ensures convergence at
threshold, and the upper limit ensures that the contribution from
infinity vanishes. These bounds on the amplitude must hold along the
whole deformed contour. In the forward limit we use $\gamma=0$ for
the massless model and $\gamma=1/2$ for the massive model. At nonzero
transfer momentum the same threshold behavior is an extra assumption. When we
differentiate with respect to $t$ or exchange an integral with the
sum over spins, we assume that all the integrals converge uniformly.

At fixed $n$ the sum over $q$ in Eq.~\eqref{eq:end-alpha-coefficients}
is a polynomial of degree $n$ in $\alpha$. A finite difference of
order $n+1$ in $\alpha$ annihilates this polynomial. For equally
spaced values $\alpha_0+ih$ with step $h$ we obtain the null
\begin{equation}
 \sum_{i=0}^{n+1}(-1)^{n+1-i}\binom{n+1}{i}
 \mathcal C_n(\alpha_0+ih,t)=0.
 \label{eq:4d-alpha-sampling}
\end{equation}
The left-hand side is a linear combination of the dispersive integrals
in the second line of Eq.~\eqref{eq:end-alpha-coefficients}. 
The same elimination expresses
the subtraction constant through dispersive integrals. The simplest
case is
\begin{equation}
 a_0(t)=8\mathcal C_2(0,t)-4\mathcal C_1(0,t)-8\mathcal C_2(1/2,t).
 \label{eq:app-subtraction-elimination}
\end{equation}
This relation is the first harmonic sum rule of
Eq.~\eqref{eq:harmonic-identities}.
Appendix~\ref{app:roy-resummation} organizes all relations between the
kernels with $\alpha=0$ and $\alpha=1/2$ into harmonics.

For reference we collect the assumptions that are used throughout the
appendices.
\begin{enumerate}
\item The amplitude is crossing symmetric and has no poles below
threshold.
\item At fixed transfer the amplitude is analytic in the energy
variable apart from the direct and crossed cuts, and it grows more
slowly than the second power of the energy. Twice-subtracted
dispersion relations therefore hold. We use the transfers
$-4m^2<t<4m^2$ for massive scattering, where $m$ is the mass of the
particle, and $-1<t\le0$, in units of the cutoff, for the massless
model.
\item Each partial wave lies in the unitarity disk at all energies on
the cut.
\item The partial-wave expansion converges on the cut at the momentum transfers
that we use. The sums over spins and the integrals over the energy
converge, and the amplitude is bounded by a power at threshold, as
specified above.
\end{enumerate}

\section{Harmonic resummation and Roy equations}
\label{app:roy-resummation}
\label{app:harmonic_resummation}

This appendix shows that the complete harmonic equations, including
the first-moment subtraction relation, are equivalent to the
real-part dispersion relation at fixed transfer. Angular projection
then gives Roy equations on the common validity domain. The logic of this construction is the following. Each
harmonic equation states that one Fourier mode of the real part on
the cut is equal to the same Fourier mode of its reconstruction from
the imaginary part. Two functions with equal Fourier modes are equal.
Thus, the sum over all the modes reconstructs the dispersion relation.
The projection of this relation on a partial wave is a Roy equation.

\paragraph{From the fractional kernels to the harmonics.}
We fix a transfer $t$ at which the dispersion relations hold. As in
Sec.~\ref{app:fractional-kernels}, $\mathcal A_t(x)$ is the even
amplitude with cuts at $|x|\ge1$, and $a_0(t)=\mathcal A_t(0)$. We
label the points of the right cut by the energy angle
$\phi=2\arccos(1/x)$, which runs from $0$ at threshold to $\pi$ at
infinite energy. We write $\mathcal A_{t,\phi}$ for the value of the
amplitude on the upper side of the cut. We introduce the function
\begin{equation}
 \zeta(x)=\frac{1-\sqrt{1-x^2}}{x},\qquad
 \zeta(x+i0)^2=e^{i\phi},\qquad
 \frac{4\,dx}{x^3}=\sin\phi\,d\phi .
 \label{app:eq:disk-variable}
\end{equation}
The expression in the square root is positive for $-1<x<1$. The function $\zeta(x)$ is
analytic in the cut plane, and it behaves as $x/2$ near the origin.
Its square is $\zeta^2=(2-x^2-2\sqrt{1-x^2})/x^2$, contains only
the kernels of Eq.~\eqref{eq:fractional} with $\alpha=0$ and
$\alpha=1/2$. For an integer $j\ge0$ the function
$4\zeta^{2j}\mathcal A_t/x^3$ is therefore a finite combination of the integrands of $\mathcal C_n(0,t)$ and
$\mathcal C_n(1/2,t)$ in Eq.~\eqref{eq:end-alpha-coefficients} with
$n\le j+1$. Its contour integral around the origin gives
\begin{equation}
 \frac8\pi\int_1^\infty\frac{dx}{x^3}
 \Imm[e^{ij\phi}\mathcal A_t(x+i0)]
 =\begin{cases}
 4a_1(t),&j=0,\\
 a_0(t),&j=1,\\
 0,&j\ge2.
 \end{cases}
 \label{app:eq:harmonic-identity}
\end{equation}
For $j=1$ this equation coincides with
Eq.~\eqref{eq:app-subtraction-elimination}. The deformation of the
contour requires that $\mathcal A_t/x^2\to0$ at infinity, that the
contributions of the endpoints of the cuts vanish, and that the
integrals in Eq.~\eqref{app:eq:harmonic-identity} converge.
Equation~\eqref{app:eq:harmonic-identity} proves the harmonic sum
rules~\eqref{eq:harmonic-identities} of the main text. The equation
with $j=1$ reconstructs the subtraction constant $a_0(t)$. The
equations with $j\ge2$ are the harmonic nulls.
\paragraph{From the harmonics to the dispersion relation.}
We now show that the complete set of
equations~\eqref{app:eq:harmonic-identity} with $j\ge1$ is equivalent
to the dispersion relation for the real part at fixed transfer. We
use
$\Imm(e^{ij\phi}\mathcal A_{t,\phi})
=\sin(j\phi)\Ree\mathcal A_{t,\phi}+\cos(j\phi)\Imm\mathcal A_{t,\phi}$
and the measure in Eq.~\eqref{app:eq:disk-variable}.
Equation~\eqref{app:eq:harmonic-identity} fixes the sine-Fourier
coefficients of $\sin\phi\,\Ree\mathcal A_{t,\phi}$ from the imaginary
part, with $a_0(t)$ entering at $j=1$. Equivalently, let $E_t(\phi)$
be the left-hand side minus the right-hand side of
Eq.~\eqref{eq:real-pv}, with transfer $t$ restored and
$y=\sec(\phi/2)$. The same identity reads
\begin{equation}
 \mathcal H_j(t)-a_0(t)\delta_{j1}
 =\frac2\pi\int_0^\pi d\phi\,\sin\phi\,\sin(j\phi)E_t(\phi).
 \label{app:eq:harmonic-residual}
\end{equation}
Thus the nulls are the sine-Fourier coefficients of
$\sin\phi\,E_t(\phi)$. To reconstruct the real part at a point $y>1$ of the cut, we
define the angle $\psi=2\arccos(1/y)$ of this point. We multiply the
$j$th equation by $q^{j-1}\sin(j\psi)$ and sum over $j\ge1$. Here
$0<q<1$ is an auxiliary parameter that makes the series converge. In
the limit $q\uparrow1$, the completeness of the functions
$\sin(j\phi)$ on the interval $0<\phi<\pi$ reconstructs the real part
at the angle $\psi$. In the term with the imaginary part we use
\begin{equation}
 \lim_{q\uparrow1}\sum_{j=1}^\infty q^{j-1}
 \sin(j\psi)\cos(j\phi)
 =\operatorname{PV}\frac{\sin\psi}{2(\cos\phi-\cos\psi)},
 \label{app:eq:harmonic-resummation-kernel}
\end{equation}
where $\operatorname{PV}$ denotes the principal value. The result is
\begin{equation}
 \Ree\mathcal A_t(y+i0)=a_0(t)
 +\frac{2y^2}{\pi}\operatorname{PV}\int_1^\infty
 \frac{\Imm\mathcal A_t(x+i0)}{x(x^2-y^2)}\,dx .
 \label{eq:resummed-dispersion}
\end{equation}
This is Eq.~\eqref{eq:real-pv} with the transfer dependence restored.
It also follows directly from the ordinary Cauchy representation,
\begin{equation}
 \mathcal A_t(\xi)=a_0(t)+\frac{2\xi^2}{\pi}\int_1^\infty
 \frac{\Imm\mathcal A_t(x+i0)}{x(x^2-\xi^2)}\,dx,
 \qquad \xi\ \hbox{off the cuts}.
 \label{app:eq:ordinary-complex-dispersion}
\end{equation}
Taking $\xi=y+i0$ uses
$(x-y-i0)^{-1}=\operatorname{PV}(x-y)^{-1}+i\pi\delta(x-y)$.
The delta function reproduces the imaginary part, while the real
part gives Eq.~\eqref{eq:resummed-dispersion}. The steps can be reversed, because the
weighted projection of Eq.~\eqref{eq:resummed-dispersion}
against $\sin\psi\,\sin(j\psi)$
returns Eq.~\eqref{app:eq:harmonic-identity}. The limit
$q\uparrow1$ exists at every point where the boundary value of the
amplitude is locally H\"older continuous, provided that the weighted
integrals converge absolutely. For less regular boundary values the
relation holds after smearing in $y$. Within this class of amplitudes
the complete set of harmonic equations and the dispersion relation at
fixed transfer are equivalent. A finite set of harmonics imposes only
finitely many Fourier modes of the dispersion relation.
\paragraph{Crossing symmetry and Roy equations.}
The dispersion relation~\eqref{eq:resummed-dispersion} contains the
subtraction function $a_0(t)$. Crossing symmetry relates the
subtraction functions at different momentum transfers and expresses them
through a common subtraction constant and absorptive integrals. We write the result in terms of the
Mandelstam variables, and we denote by $\mu$ the integration variable
along the cut. For the massless model the heavy cut starts at $s=1$,
and the relation reads
\begin{equation}
 \Ree\mathcal M(s,t)-g_0
 =\frac1\pi\operatorname{PV}\int_1^\infty d\mu\,
 \big[K(s,t;\mu)\Imm\mathcal M(\mu,t)
       +B(t;\mu)\Imm\mathcal M(\mu,0)\big],
 \label{app:eq:massless-resummed-matching}
\end{equation}
with the kernels
\begin{equation}
 \begin{aligned}
 K(s,t;\mu)&=\frac1{\mu-s}+\frac1{\mu+s+t}
                    -\frac1\mu-\frac1{\mu+t},\\
 B(t;\mu)&=\frac1{\mu-t}+\frac1{\mu+t}-\frac2\mu.
 \end{aligned}
 \label{app:eq:massless-matching-kernels}
\end{equation}
The first term in Eq.~\eqref{app:eq:massless-resummed-matching} is
the dispersion relation at fixed $t$ subtracted at $s=0$, which gives
$\mathcal M(s,t)-\mathcal M(0,t)$. The second term is the forward
dispersion relation, which gives $\mathcal M(t,0)-g_0$. Crossing
symmetry gives $\mathcal M(0,t)=\mathcal M(t,0)$. The harmonic nulls
with $j\ge2$ must therefore be imposed together with this matching of
the subtraction constant, which corresponds to the harmonic equations
with $j=1$.

For massive scattering, we set $m=1$ and denote the value of the
Mandelstam variables at the symmetric point by $c=4/3$. The same
steps give
\begin{equation}
 \begin{split}
 \Ree\mathcal M(s,t)-32\pi g_0^{\rm m}
 =\frac1\pi\operatorname{PV}\int_4^\infty d\mu\,
 \big[&\mathcal K(\mu;s,t;c)\Imm\mathcal M(\mu,t)\\
      +&\mathcal K(\mu;t,c;c)\Imm\mathcal M(\mu,c)\big],
 \end{split}
 \label{app:eq:massive-resummed-matching}
\end{equation}
with the kernel
\begin{equation}
 \mathcal K(\mu;s,t;c)=\frac1{\mu-s}
 +\frac1{\mu-4+s+t}-\frac1{\mu-c}-\frac1{\mu-4+t+c}.
 \label{app:eq:massive-matching-kernel}
\end{equation}
Here the first term gives $\mathcal M(s,t)-\mathcal M(c,t)$, the
second term gives $\mathcal M(t,c)-\mathcal M(c,c)$, and
$\mathcal M(c,c)=32\pi g_0^{\rm m}$.

Equations~\eqref{app:eq:massless-resummed-matching}
and~\eqref{app:eq:massive-resummed-matching} are dispersion relations
at fixed momentum transfer. Roy equations~\cite{r10,r5} follow from them in the
two steps. First, the imaginary parts under the integrals should be expanded
in partial waves. Second, the dependence on the external scattering
angle should be projected on a Legendre polynomial. The result expresses the
real part of a partial wave at an external energy $s$ through
integrals of the imaginary parts of all partial waves. More explicitly, let $\mathscr E(s,t)$ be the
left-hand side minus the right-hand side of
Eq.~\eqref{app:eq:massive-resummed-matching}, and set
$t_s(\chi)=-(s-4)(1-\chi)/2$. For even $J$, the projection gives
\begin{equation}
 \begin{aligned}
 \mathcal R_J(s)
 &=\frac{\rho(s)}{16\pi}\int_0^1 d\chi\,P_J(\chi)
       \mathscr E(s,t_s(\chi))\\
 &=\Ree f_J(s)-k_J(s)
   -\sum_{\ell\ {\rm even}}\operatorname{PV}\int_4^\infty
       d\mu\,K_{J\ell}(s,\mu)\Imm f_\ell(\mu)=0 .
 \end{aligned}
 \label{app:eq:roy-projection}
\end{equation}
Here $\rho(s)=\sqrt{1-4/s}$,
$k_J(s)=2\rho(s)g_0^{\rm m}\delta_{J0}$, and $K_{J\ell}$ is obtained
by projecting the two spectral terms in
Eq.~\eqref{app:eq:massive-resummed-matching} after inserting the
partial-wave expansion. Legendre orthogonality isolates $\Ree f_J$;
the crossed terms couple it to all absorptive partial waves.
 Together with
the first part of the appendix, this shows that a Roy equation is a
specific infinite linear combination of the harmonic
equations~\eqref{app:eq:harmonic-identity}. The sum over $j$ produces
the dispersion relation, crossing symmetry fixes the subtraction
function, and the angular projection is an integral over $t$ at fixed
$s$. The complete harmonic equations at every transfer required by the angular
integral, supplemented by subtraction matching and crossing, therefore
imply the Roy equations on their common domain. This statement holds
for the Roy equations whose angular projection stays within
$t>-4m^2$, which means $s<12m^2$ for the projection over the
half-interval of the scattering angle. This is the range used in
Ref.~\cite{r26}. Thus, the complete set of harmonic
nulls allows one to reproduce the dual bounds obtained from Roy
equations~\cite{r5,r26}.

\paragraph{Smeared equations.}
In practice, we impose the dispersion relations in a smeared form.
The residual $\mathscr E(s,t)$ is defined analogously for the
massless relation~\eqref{app:eq:massless-resummed-matching}. We impose the linear
equations
\begin{equation}
 \mathscr R[w]\equiv\int ds\,dt\,w(s,t)\mathscr E(s,t)=0
 \label{app:eq:smeared-matching}
\end{equation}
for weights $w(s,t)$ for which the regulated integrals and their
interchanges converge. At fixed transfer, Eq.~\eqref{app:eq:harmonic-residual}
makes the relation to energy smearing explicit:
\begin{equation}
 \sum_{j\ge1}\lambda_j[\mathcal H_j(t)-a_0(t)\delta_{j1}]
 =\frac2\pi\int_0^\pi d\phi\,\sin\phi
   \left[\sum_{j\ge1}\lambda_j\sin(j\phi)\right]E_t(\phi).
 \label{app:eq:harmonic-energy-average}
\end{equation}
Thus the coefficients $\lambda_j$ specify an energy weight; taking
$\lambda_1=0$ uses only homogeneous nulls. Finite sums and convergent
infinite sums can both be used. The latter can sometimes be evaluated
in closed form before integration, with the help of the identity
\begin{equation}
 \sum_{j=1}^\infty \sin(j\psi)\,\omega^j
 =\frac{\omega\sin\psi}{1-2\omega\cos\psi+\omega^2},\qquad |\omega|<1,
 \label{app:eq:chebyshev-resummation}
\end{equation}
where $\psi$ is the energy angle of the external point and $\omega$ is
a complex number. A specified resummed combination imposes one energy average of
real-part matching. It need not impose every harmonic separately;
only the complete Fourier reconstruction recovers the pointwise
relation. Some weights have a finite harmonic expansion.

In this paper we use two kinds of weights. A weight of the first kind is the product
of a Legendre polynomial in the cosine of the scattering angle and of
a function of the energy. It gives a Roy equation averaged over the
energy (smeared Roy equation). Its explicit form is
given in Appendix~\ref{app:massive-resummed-registry}. For such a
weight the integral over the momentum transfer covers the whole range of the
scattering angle which restricts the external energy to a finite
window. A weight of the second kind can be any other weight, for example
a weight supported on a strip of transfers. It tests the dispersion
relation at fixed transfer without an angular projection, and the
external energy is not restricted to this window. The weights that we
use for the figures are listed in
Appendices~\ref{app:massless-bounds} and~\ref{app:massive-bounds}.

\paragraph{Ranges of validity.}
Three different ranges enter this discussion. The first is the range
of momentum transfers for which the dispersion relation at fixed $t$ holds.
For massive scattering with two subtractions this range is
$-28m^2<t<4m^2$~\cite{jinmartin,martin65,martin66,toolkit,r5}. Reference~\cite{geometricCSDR}, however, discusses a larger domain, $-32m^2<t<4m^2$. Here we use the established analyticity domain (see Fig.~8 of ~\cite{geometricCSDR}), not an extension up to the Landau curve, which is not completely proven.

The second is the range in which the harmonic construction applies.
The variable $x$ defined in this appendix requires two separate cuts. At
fixed $t$ the distance from the centre of the gap between the cuts to
each branch point is $2m^2+t/2$ in the variable $s$. This distance is
positive only for $t>-4m^2$. For $t\le-4m^2$ the direct and crossed
cuts overlap, and the harmonic equations cannot be used. The harmonic
equations also require that the partial-wave expansion converges and
that the integrals converge at the endpoints of the cuts. These
properties are assumptions. They do not follow from the unitarity of
each partial wave alone.

The third is the range of external energies of a Roy equation. A Roy
equation contain an integral over the scattering angle at a fixed external
energy $s$. We denote the cosine of the scattering angle by $\chi$,
so that $t=-(s-4m^2)(1-\chi)/2$. For identical scalars, Bose symmetry
allows us to restrict the projection to $0\le\chi\le1$. The condition
$t>-28m^2$ for which the fixed-$t$ dispersion relations are proven, then gives the range $4m^2<s<60m^2$. The projection over the full interval $-1\le\chi\le1$ would allow $s<32m^2$~\cite{r5}. The implementation of Ref.~\cite{r26} uses the smaller window of
external energies $4m^2<s<12m^2$. The certified calculation below
uses $4<s<11.9$ for its Roy averages and $12<s<48$ for
fixed-transfer strips with $-4/3\le t\le0$ ($m=1$). Thus its
constraints require only the domain $-4<t<4$, also containing
the symmetric-point coupling kernels at $t=4/3$. In that dual formulation the
multipliers of the Roy equations are restricted to this window by the
convergence of the sums over spins and by the feasibility of the dual
problem. These limits concern only the external
energy, at which the amplitude is taken, while the energy integrals in the dispersion relations extend to infinite energy in all
cases. Other representations of Roy type equations may extend the range of external
energies at the price of adding the extra terms~\cite{geometricCSDR}.

In the massless model we use $-1<t\le0$. The value $t=-1$ is included
only as a limit, in which the integrals converge. The projection with
$t=-s(1-\chi)/2$ and $0\le\chi\le1$ then requires $1<s<2$. At larger
$s$ we still smear over the strip $-1<t\le0$, but this is not a
complete angular projection. These statements refer to the model
with infrared cuts neglected and not to an exact interacting massless
amplitudes.

Summarising, at any fixed allowed momentum transfer the harmonic moments are integrated over
the whole spectral cut. This does not establish a system of Roy
equations at all external energies. In fact, harmonic nulls and Roy equations are two
different samplings of the same dispersion relations set at fixed momentum transfer. 

\section{Two dimensions: finite matching and CDD boundary measures}
\label{sec:end-2d}
\label{sec:cayley}

This appendix proves the statements of the  section on two dimensions.
With the convention of Eq.~\eqref{eq:finite-nulls}, the $n$ nulls
$N_1,\ldots,N_n$ determine the exact region of the $n+1$ couplings
$\Lambda_0,\Lambda_2,\ldots,\Lambda_{2n}$. The S matrices on the
boundary of this region are products of at most $n$ CDD factors, and
generically of exactly $n$ factors.

\subsection{Coupling maps and the two-coefficient lens}
We define $x=(s-2m^2)/(2m^2)$, so that the cuts of the amplitude are
at $|x|\ge1$, and
\begin{equation}
 z=\frac{1-\sqrt{1-x^2}}{1+\sqrt{1-x^2}},\qquad
 \mathcal A=\frac{T}{4m^2}=\sqrt{1-x^2}(S-1),\qquad z=-e^{-i\theta}
 \quad\hbox{on the upper cut}.
 \label{eq:2d-conformal-map}
\end{equation}
The square root is positive for $-1<x<1$. The variable $z$ maps the
cut plane to the unit disk, and it is equal to $\zeta(x)^2$ in the
notation of Eq.~\eqref{app:eq:disk-variable}. The angle $\theta$ of
the main text is therefore related to the energy angle $\phi$ of
Eq.~\eqref{eq:harmonic-identities} by $\theta=\pi-\phi$. On the cut
we can write $S=1+if$ with $f=r+ib$. The S-matrix is analytic and bounded
in the cut plane. The amplitude $\mathcal A$ is then bounded by a multiple of
$(x-1)^{1/2}$ at threshold, and it grows at most linearly at infinity.
The contour integral $\mathcal C_1(\alpha)$ of
Eq.~\eqref{eq:end-alpha-coefficients} is therefore admissible for
$-3/2<\alpha<1/2$. The choices $\alpha=0,-1/2,-1$ give
\begin{equation}
 \Lambda_2=\avgt{(1+\cos\theta)b},\qquad
 \Lambda_2+\frac{\Lambda_0}{4}=\avgt{\sin\theta r},\qquad
 \Lambda_2+\frac{\Lambda_0}{2}=-\avgt{(1-\cos\theta)b}.
 \label{app:eq:2d-three-relations}
\end{equation}
The first and the third relations contain only the imaginary part $b$.
Their difference and their half-sum give the two relations with
$\Imm f$ in Eq.~\eqref{eq:2d-maps}. The second relation expresses the
combination $\Lambda_2+\Lambda_0/4$ through the real part $r$. The
equality of the two expressions for this combination is the null
$N_1=0$ of Eq.~\eqref{eq:N1}.

The blue and orange regions in Fig.~\ref{fig:2d} are using only the
relations for $b$. We define $\bar b=-\Lambda_0/4$, which is the
average $\avgt b$. Positivity $b\ge0$ gives $\bar b\ge0$ and
$0\le\Lambda_2\le2\bar b$, because $0\le1+\cos\theta\le2$. The
condition $0\le b\le2$ gives
\begin{equation}
 0\le\bar b\le2,\qquad
 \bar b-\frac2\pi\sin\frac{\pi\bar b}{2}
 \le\Lambda_2\le
 \bar b+\frac2\pi\sin\frac{\pi\bar b}{2}.
 \label{app:eq:2d-bounded-absorption}
\end{equation}
The two limits are reached when $b=2$ on an interval of length
$\pi\bar b/2$ at one end of the interval $0\le\theta\le\pi$ and $b=0$
elsewhere.

For the upper edge of the green region, the extremal phase of
Eq.~\eqref{eq:phase} with the multiplier $\lambda=(1-\mu^2)/2$ is
\begin{equation}
 r_*=\frac{(1-\mu^2)\sin\theta}{1+\mu^2+2\mu\cos\theta},\qquad
 b_*=\frac{(1+\mu)^2(1+\cos\theta)}{1+\mu^2+2\mu\cos\theta},
 \qquad -1<\mu<1.
 \label{eq:2d-extremal-density}
\end{equation}
Here $\mu$ is the parameter of the main text that labels the
combination $\Lambda_2+(1-\mu)\Lambda_0/4$. The averages of the
extremal phase are $\avgt{b_*}=1+\mu$ and
$\avgt{b_*\cos\theta}=\avgt{r_*\sin\theta}=(1-\mu^2)/2$. The extremal
phase therefore satisfies the null $N_1=0$ and saturates the bound of
Eq.~\eqref{app:eq:disk-support}. The reflection
$b(\theta)\mapsto b(\pi-\theta)$,
$r(\theta)\mapsto-r(\pi-\theta)$ preserves the null and gives the
lower edge. Both edges are described by
\begin{equation}
 \Lambda_0=-4(1+\mu),\qquad
 \Lambda_2=1+\mu\pm\frac{1-\mu^2}{2},\qquad
 S_\pm(z)=\frac{-\mu\pm z}{1\mp\mu z},
 \label{eq:2d-boundary-param}
\end{equation}
where the upper sign corresponds to the upper edge. At the endpoints
$\mu\to\pm1$ the S-matrix becomes a constant.

\paragraph{The Gram matrix derivation.}
\label{sec:app-2d-matrix}
Here we derive the same S-matrix constraints from a Gram matrix. We define
$q_\pm=\sqrt{1\pm\cos\theta}$. We use the four weights
$q_-(1,0)^T$, $q_-(0,1)^T$, $q_+(1,0)^T$ and $q_+(0,1)^T$ in
Eq.~\eqref{app:eq:weighted-gram} with the average $\avgt{\cdot}$.
With the relations~\eqref{app:eq:2d-three-relations} we obtain
\begin{equation}
 \begin{pmatrix}
 -\Lambda_0/2-\Lambda_2&R_-&I_s&\Lambda_2+\Lambda_0/4\\
 R_-&2+\Lambda_0/2+\Lambda_2&\Lambda_2+\Lambda_0/4&4/\pi-I_s\\
 I_s&\Lambda_2+\Lambda_0/4&\Lambda_2&R_+\\
 \Lambda_2+\Lambda_0/4&4/\pi-I_s&R_+&2-\Lambda_2
 \end{pmatrix}\succeq0,
 \quad R_\pm=\avgt{r(1\pm\cos\theta)},\quad I_s=\avgt{b\sin\theta}.
 \label{eq:4x4}
\end{equation}
The entries $\Lambda_2+\Lambda_0/4$ are the averages
$\avgt{r\sin\theta}$, which are fixed by the null $N_1=0$. The moments
$R_\pm$ and $I_s$ are not fixed by the couplings. The principal minor
formed by the rows and columns 1 and 4 is the matrix of
Eq.~\eqref{eq:2d-principal-minors}, and it does not contain these
moments. Its determinant is
$-\Lambda_0-2\Lambda_2-\Lambda_0^2/16$, and its positivity gives the
upper parabola in Eq.~\eqref{eq:lens}. The minor formed by the rows
and columns 2 and 3 has the determinant
$2\Lambda_2-\Lambda_0^2/16$, and its positivity condition gives the lower
parabola. The remaining entries of the $4\times4$ matrix are not
needed $\Lambda_0$ and $\Lambda_2$ constraints.

The obtained lens is not only an outer bound. Every point of the lens has an
analytic completion. To show this we define
$c_0=1+\Lambda_0/4$ and $c_1=2\Lambda_2+\Lambda_0/2$. The next
subsection shows that these are the first two Taylor coefficients of
$S(z)$. In these variables the lens reads $|c_0|\le1$ and
$|c_1|\le1-c_0^2$. For $|c_0|<1$ the function
\begin{equation}
 S(z)=\frac{c_0+\eta z}{1+c_0\eta z},\qquad
 \eta=\frac{c_1}{1-c_0^2}\in[-1,1]
 \label{eq:two-coefficient-completion}
\end{equation}
is analytic and bounded by one in the unit disk, and it has the
required coefficients. For $|c_0|=1$ the S-matrix is the constant
$c_0$. This proves that the lens is the exact allowed region of the two
couplings.

\subsection{The finite-null theorem}
\paragraph{Statement.}
We consider functions $S$ on the cut that satisfy $|S|\le1$ and the
$n$ nulls $N_1=\cdots=N_n=0$. The set of values that the coefficients
$c_0,\ldots,c_n$ defined in Eq.~\eqref{eq:fourier-matching} can take
for such functions is equal to the set of the first $n+1$ Taylor
coefficients of real analytic functions that are analytic and bounded
by one in the unit disk. At every boundary point of this set the
function $S$ is a product of at most $n$ CDD factors.

\paragraph{Proof.}
We start from an arbitrary measurable
function $S$ on the cut, which is an arbitrary pair of functions
$r(\theta)$ and $b(\theta)$ in the unitarity disk. We regard $S$ as a
function on the upper half of the unit circle, with $z=e^{i\varphi}$
and $\varphi=\pi-\theta$. We extend it to the lower half of the circle
by complex conjugation, which is the condition of real analyticity.
We define its Fourier coefficients
$\widehat S_k=(2\pi)^{-1}\int_{-\pi}^{\pi}S(e^{i\varphi})e^{-ik\varphi}d\varphi$.
Since $\Ree S=1-b$ and $\Imm S=r$, we find
\begin{equation}
 c_0=\widehat S_0=1-\avgt b,\qquad
 \widehat S_{-j}=(-1)^jN_j,\qquad
 c_j:=\widehat S_j+\widehat S_{-j}
       =2(-1)^{j+1}\avgt{\cos(j\theta)b}.
 \label{eq:fourier-matching}
\end{equation}
The null $N_j=0$ therefore states that the Fourier coefficient
$\widehat S_{-j}$ vanishes. A function on the unit circle is the
boundary value of a function analytic in the unit disk exactly when
all its Fourier coefficients with negative index vanish. The nulls
$N_j$ are therefore the conditions of analyticity of $S$ in the unit
disk, imposed one Fourier mode at a time. When the null $N_j$ holds,
$c_j$ is equal to the Fourier coefficient $\widehat S_j$, which is the
$j$th Taylor coefficient of an analytic function $S(z)$. The
coefficients
$c_j$ are linear combinations of the couplings
$\Lambda_0,\ldots,\Lambda_{2j}$. The first three are
\begin{equation}
 c_0=1+\frac{\Lambda_0}{4},\qquad
 c_1=2\Lambda_2+\frac{\Lambda_0}{2},\qquad
 c_2=\frac{\Lambda_0}{2}+\frac{8\Lambda_4}{3}.
 \label{eq:2d-schur-data}
\end{equation}

We now impose $N_1=\cdots=N_n=0$. We consider polynomials of degree
at most $n$ in $z$. We multiply such a polynomial by $S$, and we
project the result orthogonally on the Fourier modes $0,\ldots,n$. In
the basis $1,z,\ldots,z^n$ this operation is represented by the lower
triangular Toeplitz matrix $T_n$ in
\begin{equation}
 T_n=\begin{pmatrix}
 c_0&0&\cdots&0\\
 c_1&c_0&\cdots&0\\
 \vdots&\vdots&\ddots&\vdots\\
 c_n&c_{n-1}&\cdots&c_0
 \end{pmatrix},\qquad I-T_n^\dagger T_n\succeq0.
 \label{eq:finite-schur}
\end{equation}
The entries above the diagonal would be the coefficients
$\widehat S_{-1},\ldots,\widehat S_{-n}$, and they vanish because of
the nulls. Multiplication by a function with $|S|\le1$ does not
increase the norm, and an orthogonal projection does not increase it
either. This proves the matrix inequality in
Eq.~\eqref{eq:finite-schur}. Conversely, Schur's theorem on the
coefficient problem states that this inequality is sufficient for the
existence of a function that is analytic and bounded by one in the
unit disk and has the Taylor coefficients
$c_0,\ldots,c_n$~\cite{schur17,schur18,r14}. If $S(z)$ is such a
function, then the average of $S(z)$ and of the complex conjugate of
$S(\bar z)$ is also such a function, it has the same coefficients,
and it is real analytic. The region allowed by unitarity and by the
$n$ nulls therefore coincides with the region of the Taylor
coefficients of analytic S-matrices. This proves that the bounds are
exact.

At a boundary point of the region the matrix $I-T_n^\dagger T_n$ has
an eigenvalue equal to zero. Let $p$ be the corresponding eigenvector.
Its components in the basis $1,z,\ldots,z^n$ define a nonzero
polynomial of degree at most $n$, which we also call $p$. For this
polynomial the projection of the product $Sp$ on the Fourier modes
$0,\ldots,n$ has the same norm as $p$. Equality in the two norm inequalities requires that $|S|=1$ almost everywhere on the circle and that the product $Sp$
is a polynomial of degree at most $n$. An analytic completion has the
same matrix $T_n$ and therefore the same product $Sp$. It then
coincides with the original function $S$ almost everywhere. The S
matrix at a boundary point is therefore a ratio of two polynomials of
degree at most $n$ with unit modulus on the circle. Such a function
has the form
\begin{equation}
 S(z)=e^{i\gamma}\prod_{a=1}^{q}\frac{z-z_a}{1-\bar z_a z},
 \qquad |z_a|<1,\qquad q\le n,
 \label{app:eq:finite-cdd-degree}
\end{equation}
where $q$ is the number of CDD factors and $\gamma$ is a real phase.
Real analyticity requires
that the zeros $z_a$ are real or come in complex-conjugate pairs and
that $e^{i\gamma}=\pm1$. In the interior of the region the functions
$r$ and $b$ need not be boundary values of an analytic function. The
theorem only guarantees that an analytic S-matrix with the same
couplings exists.

\paragraph{One higher-order check.}
\label{sec:higher-coefficients}
As a check we consider $n=2$. For $|c_0|<1$ the inequality in
Eq.~\eqref{eq:finite-schur} is equivalent to the lens together with
\begin{equation}
 |c_2(1-c_0^2)+c_0c_1^2|\le(1-c_0^2)^2-c_1^2.
 \label{eq:3d-coupling-boundary}
\end{equation}
For example, $\Lambda_0=-4$ gives
$|8\Lambda_4/3-2|\le1-4(\Lambda_2-1)^2$ and
$1/2\le\Lambda_2\le3/2$. The last null cannot be omitted in general.
The function $S(e^{i\varphi})=\operatorname{sgn}(\cos n\varphi)$ has
unit modulus, satisfies the first $n-1$ nulls and has
$c_0=\cdots=c_{n-1}=0$, but it has $c_n=4/\pi$. An analytic function
bounded by one with vanishing coefficients $c_0,\ldots,c_{n-1}$
satisfies $|c_n|\le1$.
\section{Massless four-dimensional bounds}
\label{app:massless-bounds}
\label{app:massless-figure-refresh}

This appendix derives the bounds for the massless model and specifies
the constraints behind each curve in
Fig.~\ref{eft4d:fig:massless-comparison}. The cutoff is set to one,
the cuts from loops of the massless particles are omitted, and all
even spins are included. We use the variable $x=1/s$, which
runs from $0$ to $1$ along the cut, and the energy angle
$\phi=2\arccos x$ of the main text. We define $L_\ell=\ell(\ell+1)$,
which is the quantity $L$ of the main text, and
$N_\ell=L_\ell(L_\ell-8)$. The spectral average of the main text and
the maps for the couplings and the nulls are
\begin{equation}
\begin{gathered}
 \avg F=\sum_{\ell\ {\rm even}}(2\ell+1)\int_0^1xF_\ell(x)\,dx,
 \qquad f_\ell=r_\ell+ib_\ell,\\
 g_0=128\avg{r\sin\phi+b\cos\phi},\qquad
 g_2=16\avg b,\qquad g_3=16\avg{(3-2L_\ell)xb},\\
 H_j=\avg{r\sin(j\phi)+b\cos(j\phi)}=0\quad(j\ge2),\qquad
 X_4=\avg{N_\ell x^2b}=0.
\end{gathered}
\label{app:eq:massless-complete-maps}
\end{equation}
The map for $g_0$ is the harmonic
equation~\eqref{app:eq:harmonic-identity} with $j=1$ in the forward
limit. The maps for $g_2$ and $g_3$ follow from the ordinary
dispersion relation~\eqref{eq:dispersion} at second order in the
variable $z$ of the main text and from its first derivative with
respect to $t$ at $t=0$.
Table~\ref{app:tab:massless-current-curves} lists the constraints
used for each curve.

Unitarity above the cutoff and analyticity below it also imply
conditions on the partial waves at the cutoff itself. For example,
the imaginary part of each partial wave must vanish continuously for
$s\to1$. We do not impose such conditions. They could only reduce the
allowed region, so our results remain valid outer bounds. These are bounds for the heavy-cut model with light-particle loop cuts omitted. Applying them quantitatively to an interacting massless theory requires accounting for those infrared contributions. They should be understood as approximate for real physics models, as IR loops are not taken into account. This assumption reads as the extra requirement that the low energy EFT at tree level is valid until the cutoff scale. This can be imposed only approximately because unitarity below the cutoff requires non-zero imaginary parts.

\begin{table}[htb]
\centering\small
\begin{tabular}{p{0.20\textwidth}p{0.72\textwidth}}
\toprule
Curve & Constraints and status of the bound\\
\midrule
Gray, both panels & We impose $0\le b_\ell\le2$ and $X_4=0$. No constraint on the real part is used.\\[2pt]
Purple, panel (a) & We impose the unitarity disk, $X_4=0$ and all harmonic nulls $H_j=0$ with $j\ge2$. The curve is the limiting Gram ellipse of Eq.~\eqref{app:eq:gram-limit} at $\beta=1/40$, which is a closed-form bound.\\[2pt]
Orange, panel (a) & We add $X_5=0$ to the purple constraints and impose one further 4$\times$4 Gram matrix.  Its ellipse is intersected with the purple ellipse.\\[2pt]
Blue, both panels & We impose the unitarity disk and only the two nulls $H_2=X_4=0$. The curve follows from the inequality~\eqref{app:eq:massless-two-null-support} with a numerical minimization over the multipliers.\\[2pt]
Green, both panels & We add to the constraints of the blue curve the harmonic nulls at zero and nonzero transfer, the crossing nulls and the smeared dispersion relations of Sec.~\ref{app:massless-extremal}. This bound is numerical.\\
\bottomrule
\end{tabular}
\caption{The table lists the constraints used for the curves in Fig.~\ref{eft4d:fig:massless-comparison}. Each curve is an outer bound obtained from a finite number of valid inequalities. Panel (b) contains no Gram curve.}
\label{app:tab:massless-current-curves}
\end{table}

\subsection{Nonforward, fractional, and crossing equations}
\label{sec:app-massless-nonforward}
At a transfer in the range $-1\le t\le0$ we use the following
quantities. We define $A_t=1+t/2$ and $Q_t=s+t/2$, so that the
variable $z=Q_t/A_t$ of the main text has cuts at $|z|\ge1$. The
quantity $A_t$ should not be confused with the amplitude
$\mathcal A_t$. We define $P_{\ell,t}=P_\ell(1+2t/s)$, which is the
Legendre polynomial of the partial-wave
expansion~\eqref{eft4d:eq:eft-normalization}. We also define the
phase $\omega_t=e^{i\phi_t}$, where the energy angle $\phi_t$ is given
by $\cos\phi_t=2A_t^2/Q_t^2-1$ and $0\le\phi_t\le\pi$. At $t=0$ the
angle $\phi_t$ reduces to $\phi$. The harmonic
equations~\eqref{app:eq:harmonic-identity} at the transfer $t$ and the
matching of the subtraction constant in
Eq.~\eqref{app:eq:massless-resummed-matching} then read
\begin{align}
 H_j(t)&=\sum_{\ell\ {\rm even}}(2\ell+1)
 \int_1^\infty\frac{A_t^2P_{\ell,t}}{Q_t^3}
 \Imm(\omega_t^jf_\ell)\,ds=0,\qquad j\ge2,
 \label{eq:end-massless-G}\\
 G_t&=128H_1(t)+8t^2\sum_{\ell\ {\rm even}}(2\ell+1)
 \int_1^\infty\left[
 \frac{P_{\ell,t}}{sQ_t(s+t)}-\frac4{s(s^2-t^2)}
 \right]b_\ell\,ds=g_0.
 \label{app:eq:massless-reconstruction}
\end{align}
The integral in Eq.~\eqref{eq:end-massless-G} with $j=1$ defines
$H_1(t)$, which does not vanish. At $t=0$ we have $H_j(0)=H_j$. The
quantity $G_t$ is the reconstruction of $g_0$ from the spectrum at the
transfer $t$, and it must be equal to $g_0$ for every $t$. The
differences $G_t-G_0$ are therefore null constraints. At $t=-1$ each of the two
terms in the square bracket of
Eq.~\eqref{app:eq:massless-reconstruction} has a pole at $s=1$. The
two poles cancel, because $P_\ell(-1)=1$ for even $\ell$. Both
families are linear equations for the same partial waves. They hold
under the assumptions of Appendix~\ref{app:roy-resummation}.

\paragraph{Fractional nulls.}
At order $n$, any linear combination of the integrals
$\mathcal C_n(\alpha,t)$ of Eq.~\eqref{eq:end-alpha-coefficients} that
annihilates all polynomials of degree at most $n$ in $\alpha$ is a null, and
it contains no unknown coefficient. For $n+2$ equally spaced
admissible values of $\alpha$ this combination is the finite
difference of order $n+1$ in Eq.~\eqref{eq:4d-alpha-sampling}. We
call these equations fractional nulls.

\paragraph{Crossing nulls.}
The crossing nulls express the full crossing symmetry of the
low-energy expansion. At the total degree $D$ in the Mandelstam
variables we consider the dispersive expressions
\begin{equation}
 \left.\frac1{q!}\partial_t^q
 \frac{\mathcal C_n(0,t)}{(1+t/2)^{2n}}\right|_{t=0},
 \qquad n\ge1,\ q\ge0,\ 2n+q=D.
 \label{app:eq:crossing-row-family}
\end{equation}
The expression~\eqref{app:eq:crossing-row-family} is the coefficient
of $Q_t^{2n}t^q$ in the low-energy expansion of the amplitude.
Crossing symmetry requires that the terms of degree $D$ are a
combination of the monomials $\mathcal S^p\mathcal T^r$ with
$2p+3r=D$, where $p$ and $r$ are nonnegative integers. Here $\mathcal S=2Q_t^2+3t^2/2$ and
$\mathcal T=-t(Q_t^2-t^2/4)$ are the invariants $s^2+t^2+u^2$ and
$stu$ written in the variables $Q_t$ and $t$. The elimination of the
unknown coefficients of these monomials leaves linear relations
between the expressions~\eqref{app:eq:crossing-row-family}. These
relations are the crossing nulls. The first of them is $X_4$. There
are seven independent crossing nulls up to degree eight, 26 up to
degree fourteen, and 35 up to degree sixteen. We write $X_4$ and
$X_6$ for the crossing nulls of degrees four and six.

\subsection{The Gram ellipse and its limit}
\label{app:massless-matrices}
\label{sec:massless-ellipses}
At each spin the measure of Eq.~\eqref{app:eq:massless-complete-maps}
is $d\nu=x\,dx$ on the interval $0\le x\le1$. Its moments are
$\nu_k=1/(k+2)$, and the construction of
Sec.~\ref{app:hankel-gram} gives the block Hankel matrix of
Eq.~\eqref{eft4d:eq:block-hankel}. The harmonic nulls contain the
real part multiplied by $\sin(j\phi)$, which is $\sqrt{1-x^2}$ times
a polynomial in $x$. For this reason we multiply the second component
of the weights by $\sqrt{1-x^2}$. The moments of $W$ then become
\begin{equation}
 \widehat Y_k=\begin{pmatrix}
 I_k&\widehat R_k\\
 \widehat R_k&2/(k+2)-2/(k+4)-I_k+I_{k+2}
 \end{pmatrix},\quad
 \widehat R_k=\int_0^1x^{k+1}\sqrt{1-x^2}\,r_\ell\,dx.
 \label{app:eq:harmonic-block}
\end{equation}
In terms of these moments the nulls are linear. For example,
$H_2=\sum_\ell(2\ell+1)
(8\widehat R_3-4\widehat R_1+8I_4-8I_2+I_0)_\ell$.
The matrix conditions are imposed at each spin separately. They
cannot be summed over all spins directly, because the sum over spins
of the reference moments diverges.

To obtain a bound that includes all spins we introduce a weight that
makes the sum over spins converge. We choose a parameter
$0<\beta<1/12$ and define $q_\ell=1+\beta N_\ell x^2$, which is
positive for all even spins. Because of the crossing null $X_4=0$ we
have $\avg{q_\ell b}=g_2/16$. We use the weights
\begin{equation}
 \sqrt{q_\ell}\binom10,\qquad
 q_\ell^{-1/2}\binom{\cos(j\phi)}{\sin(j\phi)},
 \quad j=0,\ldots,N,
 \label{app:eq:ellipse-vectors}
\end{equation}
in the Gram matrix of Eq.~\eqref{app:eq:weighted-gram}. The entries
that mix the first weight with the others are the averages
$\avg{b\cos(j\phi)+r\sin(j\phi)}$. By
Eq.~\eqref{app:eq:massless-complete-maps} they are equal to $g_2/16$,
$g_0/128$ and $H_2,\ldots,H_N$, and the harmonics vanish. The
remaining block contains the unknown partial waves. Since
$0\preceq W\preceq2I$, this block is bounded from above by the same
block with $W$ replaced by $2I$, and positivity is preserved. This
gives
\begin{equation}
 \begin{pmatrix}g_2/16&h_N^T\\h_N&2T_N\end{pmatrix}\succeq0,
 \quad h_N=(g_2/16,g_0/128,0,\ldots,0)^T,
 \quad (T_N)_{ij}=\frac1\pi\int_0^\pi f_\beta(\phi)
 \cos[(i-j)\phi]\,d\phi.
 \label{app:eq:ellipse-gram}
\end{equation}
The matrix has the size $N+2$. The function $f_\beta(\phi)$ is the
reference weight summed over spins. To obtain it in explicit form we
bound the sum over spins from above. We write $y$ for $x^2$ and
$\gamma$ for $37\beta/39$, where $\gamma$ is not the threshold
exponent of Sec.~\ref{app:fractional-kernels}. For $\ell\ge12$ we use
$N_\ell\ge(37/39)L_\ell^2$. The function $1/(1+\gamma yL^2)$ of a
continuous variable $L$ decreases with $L$, and the ratio
$(2\ell+1)/(L_\ell-L_{\ell-2})$ is at most $25/46$ for $\ell\ge12$.
The sum over the spins $\ell\ge12$ is therefore bounded by an
integral over $L$ from $L_{10}=110$ to infinity. This gives
\begin{equation}
 \sum_{\ell\ {\rm even}}\frac{2\ell+1}{1+\beta N_\ell y}
 \le S_\beta(y):=
 \sum_{\ell=0,2,\ldots,10}\frac{2\ell+1}{1+\beta N_\ell y}
 +\frac{25}{46}\frac{\arctan[1/(110\sqrt{\gamma y})]}{\sqrt{\gamma y}},
 \quad\gamma=37\beta/39,
 \label{app:eq:ellipse-spin-majorant}
\end{equation}
We then set
$f_\beta(\phi)=\pi\sin\phi\,S_\beta((1+\cos\phi)/2)/4$. The matrix
with the entries $\cos[(i-j)\phi]$ is positive semidefinite for every
$\phi$. The replacement of the exact sum over spins by the larger
function $S_\beta$ therefore increases the reference block, and
Eq.~\eqref{app:eq:ellipse-gram} remains valid with all spins
included. The Schur complement of Eq.~\eqref{app:eq:ellipse-gram}
with respect to the block $2T_N$ gives
\begin{equation}
 (g_2/16,g_0/128)Q_N\binom{g_2/16}{g_0/128}\le g_2/8,
 \qquad Q_N=(T_N^{-1})_{\{0,1\},\{0,1\}}.
 \label{eq:massless-ellipses}
\end{equation}
Here $Q_N$ denotes the $2\times2$ block of the inverse matrix
$T_N^{-1}$ with the indices $0$ and $1$. This inequality describes an
ellipse in the $(g_0,g_2)$ plane.

The limit $N\to\infty$ can be computed in closed form. The entry
$(i,j)$ of the matrix $T_N$ depends only on the difference $i-j$. A
matrix of this kind is called a Toeplitz matrix. It is fully
determined by one function, which is $f_\beta(\phi)$ in our case,
because its entries are the Fourier coefficients of this function. A
classical theorem of Szeg\H{o} gives the inverse of such a matrix in
the limit of infinite size directly in terms of the
function~\cite{grenanderSzego,simonOPUC}. The only condition is that
the logarithm of the function is integrable. This condition holds
here, because $f_\beta(\phi)$ is positive for $0<\phi\le\pi$ and
vanishes linearly at $\phi=0$. For the block $Q_N$ we need only two
numbers, the average of $\log f_\beta$ and its first Fourier
coefficient,
\begin{equation}
 \lambda_j=\frac1\pi\int_0^\pi\log f_\beta(\phi)\cos(j\phi)\,d\phi,
 \qquad j=0,1.
 \label{app:eq:outer-function}
\end{equation}
These numbers should not be confused with the multipliers of
Sec.~\ref{app:extremal-slice}. The theorem states that the block
$Q_N$ tends to
\begin{equation}
 Q_\infty=e^{-\lambda_0}
 \begin{pmatrix}1&-\lambda_1\\-\lambda_1&1+\lambda_1^2\end{pmatrix}
 \label{app:eq:Q-limit}
\end{equation}
when $N\to\infty$. The matrices $Q_N$ increase with $N$, because each
new harmonic null strengthens the inequality. The limit therefore
gives the strongest ellipse of the family. We insert $Q_\infty$ into
Eq.~\eqref{eq:massless-ellipses} and obtain
\begin{equation}
 g_0^2+B_\beta g_0g_2+C_\beta g_2^2\le D_\beta g_2,
 \qquad
 (B_\beta,C_\beta,D_\beta)=
 \frac{(-16\lambda_1,64,2048e^{\lambda_0})}{1+\lambda_1^2}.
 \label{app:eq:gram-limit}
\end{equation}For $\beta=1/40$ this is the purple ellipse in
Fig.~\ref{eft4d:fig:massless-comparison}(a), with the coefficients
$B_\beta=4.98678154\ldots$, $C_\beta=57.02223326\ldots$ and
$D_\beta=10825.06353585\ldots$.
Equation~\eqref{eft4d:eq:ellipse-limit} of the main text quotes these
values rounded. The obtained ellipse is an analytic outer bound. It is not the
exact allowed region, because the upper bounds $W\preceq2I$ and
Eq.~\eqref{app:eq:ellipse-spin-majorant} were used.

\subsection{One additional \texorpdfstring{$4\times4$}{4x4} Gram matrix}
\label{app:massless-x5-grams}

Write $f_\ell=r_\ell+ib_\ell$, $x=1/s$,
$\phi=2\arccos x$, and $L_\ell=\ell(\ell+1)$, with
\[
 \avg{F}=\sum_{\ell\ {\rm even}}(2\ell+1)
          \int_0^1xF_\ell(x)\,dx.
\]
Besides $X_4=\avg{N_\ell x^2b}=0$, where
$N_\ell=L_\ell(L_\ell-8)$, use the degree-five crossing null
\begin{equation}
 X_5=\avg{M_\ell x^3b}=0,\qquad
 M_\ell=L_\ell(2L_\ell^2-43L_\ell+150).
 \label{app:eq:massless-x5-null}
\end{equation}
It follows by subtracting one half of the ordinary
$(n,q)=(2,1)$ dispersion row from the $(1,3)$ row,
where $q$ counts transfer derivatives.

Choose the positive weight
\begin{equation}
 q_\ell(x)=1+\frac{N_\ell x^2}{20}
                 +\frac{M_\ell x^3}{1000}.
 \label{app:eq:additional-gram-weight}
\end{equation}
Only spin two has negative added terms:
$q_2=1-0.6x^2-0.216x^3\ge0.184$.
At every other even spin, $q_\ell\ge1$.
The crossing nulls therefore give
$\avg{q_\ell b}=g_2/16$.

Take the four two-component weights
\begin{equation}
 \psi_*=\sqrt{q_\ell}\binom10,\qquad
 \psi_j=\frac1{\sqrt{q_\ell}}
          \binom{\cos(j\phi)}{\sin(j\phi)},
 \quad j=0,1,2.
 \label{app:eq:additional-gram-vectors}
\end{equation}
Their Gram matrix
$\mathsf G_{ab}=\avg{\psi_a^TW_\ell\psi_b}$ is positive.
Its first row is
$(g_2/16,g_2/16,g_0/128,H_2)$.
Thus $H_2=0$ fixes this row entirely in terms of couplings.

The remaining block still contains spectral moments.
Using $W_\ell\preceq2I$ bounds it from above by twice
the reference matrix with entries
$\avg{\cos[(i-j)\phi]/q_\ell}$.
Increasing this block preserves positivity.

To make the reference constants explicit, including all spins,
define
\begin{equation}
 \begin{aligned}
 \Pi(v)&=1+\frac{37}{780}v^2
             +\frac{7019}{4056000}v^3,\\
 S(x)&=\sum_{\ell=0,2,\ldots,10}
          \frac{2\ell+1}{q_\ell(x)}
       +\frac{25}{46x}
          \int_{110x}^{\infty}\frac{dv}{\Pi(v)}.
 \end{aligned}
 \label{app:eq:additional-gram-reference}
\end{equation}
For $\ell\ge12$,
$N_\ell\ge37L_\ell^2/39$ and
$M_\ell\ge7019L_\ell^3/4056$.
A decreasing-integrand estimate of the remaining spin sum gives
\[
 \sum_{\ell\ {\rm even}}\frac{2\ell+1}{q_\ell(x)}
 \le S(x).
\]
Consequently we may use the reference block
\begin{equation}
 T=\begin{pmatrix}
 \mu_0&\mu_1&\mu_2\\
 \mu_1&\mu_0&\mu_1\\
 \mu_2&\mu_1&\mu_0
 \end{pmatrix},\qquad
 \mu_j=\int_0^1xS(x)\cos(j\phi)\,dx,
 \quad j=0,1,2.
 \label{app:eq:additional-gram-constants}
\end{equation}
These convergent integrals define the constants exactly.
Since $\cos(j\phi)$ is a polynomial in $x^2$,
exchanging the integration order reduces them to finite
rational integrals, hence elementary constants. Numerically,
\[
 \begin{aligned}
 \mu_0&=7.09817132105154\ldots,\\
 \mu_1&=0.43303994447188\ldots,\\
 \mu_2&=-0.93491046083740\ldots.
 \end{aligned}
\]
This yields Eq.~\eqref{eft4d:eq:additional-gram}.

Since $T$ is positive definite, the matrix condition is
equivalent to $g_2\ge0$ and
\begin{equation}
 \begin{aligned}
 g_0^2
 &-\frac{16\mu_1}{\mu_0+\mu_2}\,g_0g_2
 +\frac{64(\mu_0^2-\mu_1^2)}
        {\mu_0^2-\mu_2^2}\,g_2^2\\
 &\le2048\left(
       \mu_0-\frac{2\mu_1^2}{\mu_0+\mu_2}
       \right)g_2.
 \end{aligned}
 \label{app:eq:additional-gram-ellipse}
\end{equation}
With coefficients rounded for display, this reads
\[
 g_0^2-1.124184\,g_0g_2+64.887461\,g_2^2
 \lesssim14412.429824\,g_2.
\]
The orange curve intersects this inequality with the original
$N\to\infty$ ellipse. The separate linearized-unitarity cap
is not imposed on either Gram curve.

\subsection{Bounds from extremal measures}
\label{app:massless-extremal}
The blue curves in Fig.~\ref{eft4d:fig:massless-comparison} use the
two nulls $H_2$ and $X_4$. We bound the combination
$n_0g_0/128+n_2g_2/16+n_3g_3/16$, where the real numbers $n_0$, $n_2$
and $n_3$ define a direction in the space of couplings. We add the
nulls with multipliers in the form $\lambda H_2-\eta X_4$, where
$\lambda$ is real and $\eta>0$. The kernels of
Eq.~\eqref{app:eq:disk-support} and the resulting inequality are
\begin{equation}
 \begin{aligned}
 U_\ell&=n_0\sin\phi+\lambda\sin2\phi,\\
 V_\ell&=n_2+n_0\cos\phi+n_3(3-2L_\ell)x
                   +\lambda\cos2\phi-\eta N_\ell x^2,\\
 n_0g_0/128+n_2g_2/16+n_3g_3/16
 &\le\avg{V_\ell+\sqrt{U_\ell^2+V_\ell^2}}.
 \end{aligned}
 \label{app:eq:massless-two-null-support}
\end{equation}
The inequality holds for every choice of the multipliers, and we
minimize its right-hand side over $\lambda$ and $\eta$ numerically.
Panel (a) uses $n_3=0$, and panel (b) uses $n_0=0$.

The gray curves use only $0\le b_\ell\le2$ and $X_4=0$. In this case
the maximum of $V_\ell b_\ell$ at each energy is $2[V_\ell]_+$, where
$[z]_+=\max(z,0)$. With $n_0=0$ and without the null $H_2$ this gives
the bound
$n_2g_2/16+n_3g_3/16\le
2\avg{[n_2+n_3(3-2L_\ell)x-\eta N_\ell x^2]_+}$
for every real $\eta$. The gray curve in panel (b) is the
intersection of 260 inequalities of this form with the two positivity
bounds $g_3\le3g_2$ and $g_3\ge-C_Pg_2$, where
$C_P=(9+\sqrt{2989/20})/2$. In panel (a) the gray region is the strip
$0\le g_2\le125.339\ldots$, and $g_0$ is not bounded.

The sum over spins in Eq.~\eqref{app:eq:massless-two-null-support}
is evaluated explicitly up to a maximal spin, and the remaining spins
are bounded analytically. Let $E$ be the first even spin that is not
evaluated explicitly, and let $L_E=E(E+1)$. We define the auxiliary
constants
\begin{equation}
 \begin{gathered}
 k=1-8/L_E,\quad
 k'=k-\frac{4|n_0|+16|\lambda|}{\eta L_E^2}>0,\quad
 t_0=n_2-n_0+\lambda,\quad d=-2n_3+3|n_3|/L_E,\\
 J=2\int_0^\infty y[t_0+dy-\eta k'y^2/2]_+\,dy,\qquad
 A=n_0^2-4n_0\lambda/3+4\lambda^2/3.
 \end{gathered}
 \label{app:eq:massless-tail-constants}
\end{equation}
These constants are used only in the present estimate. The
contribution of all spins $\ell\ge E$ to the right-hand side of
Eq.~\eqref{app:eq:massless-two-null-support} is at most
\begin{equation}
 \left(J+\frac{A}{\eta k}\right)
 \frac{\psi_1(E/2)-\psi_1((E+1)/2)}4 ,
 \label{app:eq:massless-two-null-tail}
\end{equation}
where $\psi_1$ is the trigamma function. To derive this estimate we
write $V_\ell=(V_\ell+D)-D$ with $D=\eta N_\ell x^2/2$, and we use
the inequality
$V_\ell+\sqrt{U_\ell^2+V_\ell^2}\le2[V_\ell+D]_++U_\ell^2/(2D)$,
which holds for $D>0$. The first term leads to the integral $J$. The
second term uses $\int_0^1U_\ell^2dx/x=A$, and the remaining sum over
spins is the combination of trigamma functions in
Eq.~\eqref{app:eq:massless-two-null-tail}. The blue curves in panels
(a) and (b) are the intersections of 385 and 579 inequalities of the
form~\eqref{app:eq:massless-two-null-support}, with the spins up to
512 and 1024 evaluated explicitly. The resulting ranges of the
couplings are $-606.389\lesssim g_0\lesssim491.048$,
$g_2\lesssim124.527$, and $-735.809\lesssim g_3\lesssim30.701$.

\paragraph{Harmonic nulls and smeared dispersion relations.}
The green curves in Fig.~\ref{eft4d:fig:massless-comparison} add
further null constraints and smeared dispersion relations. We recall that
$\mathscr E(s,t)$ denotes the left-hand side minus the right-hand
side of the dispersion
relation~\eqref{app:eq:massless-resummed-matching}, including the
subtraction constant $g_0$. The smeared equations are integrals of
$\mathscr E(s,t)$ with a weight, as in
Eq.~\eqref{app:eq:smeared-matching}. We use the following five
families of constraints. The families overlap, so the numbers of
equations quoted below are not the numbers of independent constraints.
The families describe a union of tests and not a single joint
optimization.
\begin{enumerate}
\item The first family consists of harmonic and crossing nulls. It
contains the forward harmonic nulls $H_j(0)$ with $2\le j\le1552$,
the harmonic nulls $H_j(-i/16)$ with $2\le j\le1536$ at the sixteen
transfers with $1\le i\le16$, the sixteen nulls $G_{-i/16}-G_0$, the
26 crossing nulls up to degree fourteen, and the resummed forward
null $\sum_{j\ge2}(9/10)^{j-2}H_j(0)=0$. In total, these are 26,154
equations. With forward constraints only, namely
$H_2,\ldots,H_{48}$, $X_4$, $X_6$ and the 94 fractional nulls given
by the third differences of $\mathcal C_2(\alpha,0)$ at
$\alpha=i/128$ with $0\le i\le96$, we obtain $g_2\lesssim116.186$.

\item The second family consists of the smeared Roy equations applied to the massless case with the forced analyticity gap, which we
use to improve the bounds on $g_0$. We impose
\begin{equation}
 2\int_1^B ds\int_0^1d\chi\,
 \frac{P_n(2(s-1)/(B-1)-1)}{B-1}P_J(\chi)
 \mathscr E(s,-s(1-\chi)/2)=0,
 \quad B=1999/1000,
 \label{app:eq:massless-angular-tests}
\end{equation}
with $n=0,\ldots,95$ and $J=0,2,4,6,8$. Here $\chi$ is the cosine of
the scattering angle, and $B$ is the upper end of the window of
external energies. These are 480 equations. The equation with
$n=0$ and $J=0$ contains the subtraction constant $g_0$, and the
other 479 equations are null constraints. The crossing null $X_4=0$ is imposed
together with them. The index $J$ labels the test function and is
not a restriction on the spins of the partial waves. The external
transfers lie in the interval $-0.9995\le t\le0$, and the spectral
integral inside $\mathscr E$ extends to infinite energy. The plotted
inequalities include those obtained with the subset $n\le31$ of
these equations. They also include inequalities whose multipliers are
combinations of the multipliers of retained inequalities, with the
bound evaluated again.

\item The third family consists of dispersion relations smeared in
the energy and in the momentum transfer, which we use for the boundary in the
$(g_2,g_3)$ plane. We use
Eq.~\eqref{app:eq:massless-angular-tests} with $n=0,\ldots,15$, and
we add the equations
\begin{equation}
 \int_A^B ds\int_{-1}^0dt\,
 \frac{6y(1-y)}{B-A}P_n(2y-1)
 30v^2(1-v)^2P_k(2v-1)\mathscr E(s,t)=0,
 \quad y=\frac{s-A}{B-A},\ v=-t,
 \label{app:eq:massless-strip-tests}
\end{equation}
for the energy windows $[A,B]=[1,4],[2,8],[4,16],[8,32]$, with
$n=0,\ldots,7$ and $k=0,\ldots,5$. In this equation $A$ and $B$
denote the ends of the energy window. There are $80+192=272$
equations, and $X_4=0$ is imposed together with them. We retain 29
inequalities evaluated with these equations.

\item The fourth family consists of the resummed harmonic nulls at
nonzero momentum transfer, which we use for the upper bound on $g_2$. These
nulls are convergent infinite combinations of the harmonic nulls at
the sixteen values $t_i=-(2i+1)/32$ with $i=0,\ldots,15$. For
$p=1,2$, $r=13/20,9/10$, and
$(\delta,\sigma)=(0,c),(\pi/2,c),(\pi/2,s),(\pi,c)$ we define
\begin{equation}
 A_\pm(\omega)=\frac{\omega^2}{(1-re^{\pm i\delta}\omega)^p},
 \quad S_c=\frac{A_++A_-}{2},\quad S_s=\frac{A_+-A_-}{2i},
 \quad F(\omega)=(1-r)^p[S_\sigma(\omega)-S_\sigma(-1)\omega^2].
 \label{app:eq:massless-rational-nulls}
\end{equation}
In this equation, $p$ is an integer power, $r$ is a real-valued parameter, $\delta$ is an angle, and the label $\sigma$ takes
the values $c$ and $s$. The null constraints are obtained from
Eq.~\eqref{eq:end-massless-G} by the replacement of $\omega_t^j$ with
$F(\omega_t)$. This defines 256 nulls. Each of them is a convergent
series of the harmonic nulls $H_j(t_i)$ with $j\ge2$, because
$F(\omega)$ is a power series in $\omega$ that starts at $\omega^2$.
The term proportional to $S_\sigma(-1)\omega^2$ removes the leading
behavior of the kernel at high energy, where $\omega_t\to-1$.
Being added to a bound that uses the crossing nulls up to degree fourteen,
these null constraints give $g_2\le68.3867533$.

\item The fifth family consists of higher crossing nulls, which we
use for the upper bound on $g_3$. We add the 50 crossing nulls of
degrees 9 to 20. This gives $g_3\le17.94753$. The inclusion of all
148 crossing nulls of degrees 9 to 32 does not change this value,
because the multipliers of the nulls above degree twenty are below
$10^{-42}$.
\end{enumerate}
We also tested the fractional nulls of
Eq.~\eqref{eq:4d-alpha-sampling}. We used 45 finite differences with
$n=2,3,4,6,8$ at the transfers $t=0,-1/2,-1$, with $n+2$ equally
spaced values of $\alpha$ in each of the intervals $[0,3/4]$,
$[0,n-21/20]$ and $[-2/5,n-21/20]$. These nulls and the 94 fractional
nulls of the forward test in the first family do not change the green
region. Its boundary is determined by the harmonic nulls, by the
crossing nulls, and by the smeared dispersion relations.

The resulting ranges of the couplings for the green region are
$-173.32\lesssim g_0\lesssim24.85$, $0\le g_2\lesssim68.387$, and
$-338.25\lesssim g_3\lesssim17.948$. All spins are included in these
bounds. The low spins are integrated numerically, and the remaining
spins are bounded analytically. The explicit integration
extends to spin 128 for the bounds on $g_0$, to spin 1024 for the
boundary in the $(g_2,g_3)$ plane, and to spin 2048 for the upper
bound on $g_2$. The integrals over the explicitly treated spins are
evaluated by numerical quadrature and not by interval arithmetic.
These bounds are therefore numerical, and we do not claim that the
boundary is reached by an amplitude. The directions, the multipliers
and the reference integrals that define the plotted curves are
provided in the accompanying data.

\paragraph{Comparison with the primal bounds of Ref.~\cite{r12}.}
Reference~\cite{r12} obtained primal bounds for EFT amplitudes of
this type. A direct comparison with our bounds is difficult for two
reasons. First, the convergence of the primal result of
Ref.~\cite{r12} with respect to the discretization in the energy was
not rigorously studied. Second, Ref.~\cite{r12} effectively considers only
dispersion relations without subtractions, whereas we study
twice-subtracted dispersion relations, and this makes our bounds
weaker.

\section{Massive four-dimensional bounds}
\label{app:massive-bounds}
\label{app:massive-matrices}

This appendix derives the massive Gram bounds and the small-null
example, then gives the constraints and certificates for
Fig.~\ref{eft4d:fig:massive-comparison}. We use the normalization and
spectral average of Eq.~\eqref{eft4d:eq:main-massive-measure}, with
$m=1$, and write $\bar s=s-4/3$, $b_\ell=\Imm f_\ell$ and
$r_\ell=\Ree f_\ell$. Higher couplings remain free.
Table~\ref{app:tab:certified-banks} lists the two constraint sets
used for the certified enclosure.

\subsection{Coupling maps and null equations}
\label{app:massive-symmetric-identities}
At the transfer $t=4/3$ the variable $z=3\bar s/8$ of the main text
has cuts at $|z|\ge1$, and the symmetric point is $z=0$. The constant
and the quadratic coefficient of $\mathcal M/(32\pi)$ in the variable
$z$ are $g_0^{\rm m}$ and $128g_2^{\rm m}/9$. The harmonic
equations~\eqref{app:eq:harmonic-identity} then give
Eq.~\eqref{eft4d:eq:main-massive-maps}. More generally, the ordinary
dispersion relation~\eqref{eq:dispersion} at higher orders gives
\begin{equation}
 I_k^*=\left\langle\frac{p_\ell}{\bar s^{2k}}b_\ell\right\rangle_{\rm m}
       =2^k g_{2k+2}^{\rm m},\qquad
 p_\ell=\frac{P_\ell(\bar s/(s-4))}{\bar s^3}>0.
 \label{app:eq:massive-positive-moments}
\end{equation}
Here $g_{2n}^{\rm m}$ is the coefficient of
$(\bar s^2+\bar t^2+\bar u^2)^n$ in the expansion of
Eq.~\eqref{eft4d:eq:massive-coefficients}. The other crossing
invariant $\bar s\bar t\bar u$ vanishes at $t=4/3$. The weight
$p_\ell$ is positive, because the argument of the Legendre polynomial
is larger than one. Positivity of $b_\ell$ alone therefore gives the
Hankel condition
\begin{equation}
 \begin{pmatrix}g_2^{\rm m}&2g_4^{\rm m}\\
 2g_4^{\rm m}&4g_6^{\rm m}\end{pmatrix}\succeq0,
 \qquad (g_4^{\rm m})^2\le g_2^{\rm m}g_6^{\rm m}.
 \label{eft4d:eq:massive-absorptive-hankel}
\end{equation}
This condition does not involve the real part and does not bound
$g_0^{\rm m}$.

In the forward limit $t=0$ the even variable is $z=(s-2)/2$. The
corresponding energy angle $\vartheta_{\rm m}$ is defined by
$\cos\vartheta_{\rm m}=8/(s-2)^2-1$, and we define the weight
$F=32/(s-2)^3$. The forward harmonic constraints and the reconstruction of the
constant coupling are
\begin{equation}
 \begin{gathered}
 \widetilde H_1(0)=\langle F\Imm(e^{i\vartheta_{\rm m}}f)\rangle_{\rm m},\qquad
 \widetilde H_j(0)=\langle F\Imm(e^{ij\vartheta_{\rm m}}f)\rangle_{\rm m}=0
 \quad(j\ge2),\\
 g_0^{\rm m}=\widetilde H_1(0)+\langle C_\ell b_\ell\rangle_{\rm m},\qquad
 C_\ell=\frac8{(3s-8)(3s-4)}
 \left[\frac1{s-2}-\frac4s P_\ell\!\left(\frac{\bar s}{s-4}\right)\right].
 \end{gathered}
 \label{app:eq:massive-forward-map}
\end{equation}
The first forward harmonic $\widetilde H_1(0)$ reconstructs the
forward amplitude at $s=2$, which differs from the amplitude at the
symmetric point. The term with $C_\ell$ accounts for this difference
through the matching of the subtraction constants in
Eq.~\eqref{app:eq:massive-resummed-matching}. The forward moment
$\widetilde H_1(0)$ is therefore different from the symmetric-point
moment $9g_0^{\rm m}/512$ of
Eq.~\eqref{eft4d:eq:main-massive-maps}. Finite sets of forward nulls
and of symmetric-point nulls are different truncations of the same
dispersion relations.

At a general transfer $t$ the harmonic nulls and the reconstruction
of $g_0^{\rm m}$ read
\begin{equation}
 \begin{aligned}
 \widetilde H_j(t)&=\langle F_tP_{\ell,t}\Imm(\omega_t^j f_\ell)\rangle_{\rm m}=0,
 \quad j\ge2,\\
 G_t&=\langle F_tP_{\ell,t}\Imm(\omega_t f_\ell)
 -(P_{\ell,t}K_A+P_{\ell,*}K_B)b_\ell\rangle_{\rm m}=g_0^{\rm m},
 \end{aligned}
 \label{eq:end-massive-G}
\end{equation}
with the definitions
\begin{equation}
 \begin{gathered}
 Q_t=s-2+t/2,\quad A_t=2+t/2,\quad F_t=8A_t^2/Q_t^3,\\
 P_{\ell,t}=P_\ell\!\left(1+\frac{2t}{s-4}\right),\quad
 P_{\ell,*}=P_\ell(\bar s/(s-4)),\quad
 \omega_t=\frac{(A_t+i\sqrt{(s-4)(s+t)})^2}{Q_t^2},\\
 K_A=\frac2{Q_t}-\frac1{\bar s}-\frac1{s-8/3+t},\qquad
 K_B=\frac1{s-t}+\frac1{s-8/3+t}-\frac2{\bar s}.
 \end{gathered}
 \label{app:eq:massive-transfer-kernels}
\end{equation}
These definitions are the massive analogues of those in
Sec.~\ref{sec:app-massless-nonforward}. The even variable is
$z=Q_t/A_t$, and the phase $\omega_t=e^{i\phi_t}$ satisfies
$\cos\phi_t=2A_t^2/Q_t^2-1$. The polynomial $P_{\ell,*}$ is the
Legendre polynomial at the transfer $4/3$. The kernel $K_A$
multiplies the partial-wave expansion at the transfer $t$, and the
kernel $K_B$ multiplies the expansion at the transfer $4/3$, in
correspondence with the two terms of
Eq.~\eqref{app:eq:massive-resummed-matching}. The first term of $G_t$
defines $\widetilde H_1(t)$. At $t=0$ these expressions reduce to
Eq.~\eqref{app:eq:massive-forward-map}. In particular,
$G_{4/3}=G_0$ and $\widetilde H_j(4/3)=(512/9)\mathcal H_j^*$, where
$\mathcal H_j^*$ are the symmetric-point harmonics of the main text.
The equations hold for transfers $t>-4$, for which the two cuts do
not overlap, and under the assumptions of
Appendix~\ref{app:roy-resummation}. In the forward limit the
fractional kernels are admissible for $-1/2<\alpha\le n-1$.
Derivatives with respect to $t$ require that the integrals converge
locally uniformly in $t$.

The first crossing null at the symmetric point is
\begin{equation}
 \widetilde X_4=\left\langle
 \left[\frac{2P_\ell''(Z)}{\bar s^3(s-4)^2}
 -\frac{3P_\ell'(Z)}{\bar s^4(s-4)}\right]b_\ell\right\rangle_{\rm m}=0,
 \qquad Z=\bar s/(s-4),
 \label{eft4d:eq:massive-crossing}
\end{equation}
where the primes denote derivatives of the Legendre polynomial with
respect to its argument. The higher crossing nulls are constructed as
in Sec.~\ref{sec:app-massless-nonforward}. In
Eq.~\eqref{app:eq:crossing-row-family} the factor $(1+t/2)^{2n}$ is
replaced by $(2+t/2)^{2n}$, and the derivatives are taken at $t=4/3$.
The crossing-symmetric monomials are built from
$2Q_t^2+3(t-4/3)^2/2$ and $-(t-4/3)[Q_t^2-(t-4/3)^2/4]$. There are
seven independent crossing nulls up to degree eight and 35 up to
degree sixteen.

We also use nulls that contain derivatives with respect to $t$ at
$t=0$. Each derivative with respect to $t$ makes the kernels more
singular at threshold. We therefore differentiate combinations of
harmonics whose kernels vanish at threshold to a sufficient order. We
define
\begin{equation}
 \begin{aligned}
 \mathcal Q_2&=\partial_t^2(G_t-2\widetilde H_2+\widetilde H_3)|_0,\\
 \mathcal Q_3&=\partial_t^3(G_t-4\widetilde H_2+6\widetilde H_3
                  -4\widetilde H_4+\widetilde H_5)|_0,\\
 \mathcal D_{j,p}&=\left.\partial_t^p
 \sum_{q=0}^{2p-2}(-1)^q\binom{2p-2}{q}
 \widetilde H_{j+q}(t)\right|_0,\qquad p=2,3.
 \end{aligned}
 \label{eq:app-balanced-derivatives}
\end{equation}
In these expressions $\widetilde H_j$ stands for $\widetilde H_j(t)$,
and the derivatives are evaluated at $t=0$. All these quantities
vanish, because $G_t$ does not depend on $t$ and the harmonics with
$j\ge2$ vanish for every $t$. The alternating sums are finite
differences in the index $j$. They multiply the harmonic kernels by
the factor $(1-\omega_t)^{2p-2}$, which vanishes at threshold, where
$\omega_t=1$. This factor compensates the threshold singularities
produced by the derivatives.

\subsection{Bounds from extremal measures: Two nulls}
\label{app:massive-two-null}
With the two nulls $\widetilde H_2(0)$ and $\widetilde X_4$ only, we
use the kernels of $g_0^{\rm m}$, $g_2^{\rm m}$, $\widetilde H_2(0)$
and $\widetilde X_4$ in the inequality of
Eq.~\eqref{app:eq:disk-support}. The numerical value of the upper
bound is $g_0^{\rm m}\simeq4.343$. With these two nulls the coupling
$g_2^{\rm m}$ is not bounded from above. This is shown by the
following family of partial waves. We set $r_\ell=0$ for all spins,
and in the interval $\epsilon<s-4<2\epsilon$ we choose
\begin{equation}
 b_2=2,\qquad
 b_4=\frac{4(s-4)^2}{9[7(s-4/3)^2+(s-4)^2]},
 \label{app:eq:threshold-example}
\end{equation}
which makes the integrand of $\widetilde X_4$ vanish at every energy.
A contribution of the imaginary part of the spin-zero partial wave
of order $\sqrt\epsilon$ near $s=5$ cancels the remaining harmonic
$\widetilde H_2(0)$. In the limit $\epsilon\to0$ the coupling
$g_2^{\rm m}$ grows as $\epsilon^{-3/2}$, and the ratio
$g_0^{\rm m}/g_2^{\rm m}$ tends to $-128/27$.

\subsection{Certified bounds from extremal measures}
\label{app:massive-resummed-registry}

The region shown in Fig.~\ref{eft4d:fig:massive-comparison} is
the intersection of 25 certified supporting inequalities and
$g_2^{\rm m}\ge0$. We retain inequalities that improve the published
dual comparison, according to the criterion below. We describe the equations used to find these
inequalities, then explain how their integrals and remainders are
bounded. The calculation uses the fixed-transfer assumptions stated
above, including the convergence needed for differentiation and
smearing. 

\paragraph{The constraint sets.}
Two finite sets of equations are used, listed in
Table~\ref{app:tab:certified-banks}. Seven retained inequalities use set I; eighteen use set II,
including positive combinations of witnesses from that set. The row counts specify the retained equations,
not a claim that they are independent or exhaustive. Higher EFT
coefficients remain free. The two sets contain neither additional
fractional-$\alpha$ samples nor finite-point crossing tests.

\begin{table}[t]
\centering\small
\begin{tabular}{p{.23\textwidth}p{.34\textwidth}p{.34\textwidth}}
\toprule
Equations & Set I: 150 rows & Set II: 220 rows\\
\midrule
Forward harmonics & $\widetilde H_j(0)$, $j=2,\ldots,40$ (39) & Same (39)\\
Resummed harmonics & 14 cosine sums & 7 cosine and 7 sine sums\\
Crossing & $\widetilde X_4$ (1) & Seven nulls through degree eight\\
Smeared Roy & $J=0,2,4,6$; $n=0,\ldots,14$ (60) & Same (60)\\
Fixed-transfer strips & Two energy windows; $j=0,1,2$, $n=0,\ldots,5$ (36) & Same windows; $j=0,\ldots,3$, $n=0,\ldots,7$ (64)\\
Transfer derivatives & None & Orders $p=1,2,3$ (36)\\
\bottomrule
\end{tabular}
\caption{The complete equation families used by the 25 retained
numerical inequalities. Their parameters and normalizations are given below.
Both sets use method A of the main text.}
\label{app:tab:certified-banks}
\end{table}

For clarity, we denote the spectral energy by $\mu$ in this
subsection. It always ranges over $4<\mu<\infty$. The forward
kernels $F$, $C_\ell$, and $\omega=e^{i\vartheta_{\rm m}}$ are those
of Eq.~\eqref{app:eq:massive-forward-map}, with $s$ replaced by $\mu$.
Thus the two coupling maps used in the calculation are
\begin{equation}
 g_0^{\rm m}=\langle u_0r+v_0b\rangle_{\rm m},\qquad
 g_2^{\rm m}=\langle p_\ell b_\ell\rangle_{\rm m},\qquad
 u_0=F\Imm\omega,\quad v_0=F\Ree\omega+C_\ell.
 \label{app:eq:certified-coupling-maps}
\end{equation}
The crossed subtraction term $C_\ell$ is retained. The numerical
calculation therefore bounds the symmetric-point coupling, not the
forward amplitude alone.

\paragraph{Forward harmonics and their sums.}
Both sets contain $\widetilde H_2(0),\ldots,\widetilde H_{40}(0)$.
They also contain the exact rational kernels
\begin{equation}
 \begin{gathered}
 q^{c/s}_{r,\zeta}=F\omega^2[m_{c/s}(\omega)-m_{c/s}(-1)],
 \qquad \langle\Imm(q^{c/s}_{r,\zeta}f)\rangle_{\rm m}=0,\\
 m_c(w)=\frac{1-r\zeta w}{1-2r\zeta w+r^2w^2},\qquad
 m_s(w)=\frac{rw}{1-2r\zeta w+r^2w^2}.
 \label{app:eq:massive-rational-bank}
 \end{gathered}
\end{equation}
Here $r<1$ is a summation parameter and $\zeta$ is real. The power
series of $m_c$ and $m_s$ have coefficients $r^kT_k(\zeta)$ and
$r^kU_{k-1}(\zeta)$, respectively. Each equation is consequently one
convergent combination of harmonic nulls. It does not impose every
harmonic in that combination separately. The subtraction at
$\omega=-1$ improves the high-energy decay.
Set I uses only $m_c$, with
\begin{equation}
 r\in\{4/5,19/20\},\qquad
 \zeta\in\{-1,-3/4,-1/2,0,1/2,3/4,1\}.
\end{equation}
Set II uses both functions at $r=4/5$, with
\begin{equation}
 \zeta\in\{-9/10,-3/4,-1/2,0,1/2,3/4,9/10\}.
\end{equation}
All combinations of the stated parameters are included.

\paragraph{Crossing and transfer derivatives.}
The crossing equations are evaluated at $s=t=u=4/3$. To fix their
normalization, write $Q=\mu-4/3$ and $Z=Q/(\mu-4)$:
\begin{equation}
 \widetilde X_{D,q}=
 \left\langle\frac{b_\ell}{Q^{D+1}}
       \sum_{j=1}^{D-2}a_{D,q,j}Z^jP_\ell^{(j)}(Z)
 \right\rangle_{\rm m}=0.
 \label{app:eq:certified-crossing}
\end{equation}
The derivatives here are ordinary derivatives, without division by
$j!$. Set I uses the first row of the following table, and set II
uses all seven:
\begin{equation}
\begin{array}{c|rrrrrr}
(D,q)&a_1&a_2&a_3&a_4&a_5&a_6\\\hline
4&-3&2&0&0&0&0\\
5&8&-12&16/3&0&0&0\\
6&-160&128&-64&64/3&0&0\\
7,1&30&-70&72&-40&32/3&0\\
7,2&48&-80&64&-32&128/15&0\\
8,1&112&-144&160/3&32/3&-64/5&128/45\\
8,2&-784&720&-1120/3&352/3&-128/5&256/45
\end{array}
\end{equation}
These rows follow by eliminating crossing-symmetric coefficients
at each EFT degree. In particular, the first row is exactly
Eq.~\eqref{eft4d:eq:massive-crossing}.

The 36 transfer equations in set II are
\begin{equation}
 G'_0,\quad\mathcal Q_2,\quad\mathcal Q_3,\quad
 \widetilde H'_j(0),\quad\mathcal D_{j,2},\quad\mathcal D_{j,3},
 \qquad j=2,\ldots,12.
 \label{app:eq:certified-transfer-rows}
\end{equation}
The combinations $\mathcal Q_p$ and $\mathcal D_{j,p}$ are defined in
Eq.~\eqref{eq:app-balanced-derivatives}. Their threshold zeros are
inserted before differentiation. These zeros reduce the threshold singularity introduced by
the derivatives; the convergence assumptions stated above still
apply.

\paragraph{Roy equations and fixed-transfer strips.}
Both families smear the same residual $\mathscr E(s,t)$ of
Eq.~\eqref{app:eq:massive-resummed-matching}. The variable $s$ below
is the external energy where that relation is tested. Its integration
window does not restrict the spectral energy $\mu$.
The retained Roy equations are
\begin{equation}
 \begin{gathered}
 \mathcal R_{Jn}=\frac1{32\pi}\int_4^B ds\,w_{Bn}(s)
       \int_0^1 2\,d\chi\,P_J(\chi)
       \mathscr E\!\left(s,-\frac{s-4}{2}(1-\chi)\right)=0,\\
 w_{Bn}(s)=\frac1{B-4}
       P_n\!\left(\frac{2(s-4)}{B-4}-1\right),\qquad
 B=\frac{119}{10},\quad J=0,2,4,6,\quad n=0,\ldots,14.
 \label{app:eq:massive-angular-registry}
\end{gathered}
\end{equation}
The projection covers the half-angle interval allowed by Bose
symmetry. Its transfers satisfy $-3.95\le t\le0$.
Only the row $J=n=0$ contains an explicit subtraction constant,
with coefficient $-2g_0^{\rm m}$. Substituting
Eq.~\eqref{app:eq:certified-coupling-maps} makes it a homogeneous
spectral identity, as required for adding it to a support functional.

To test higher external energies, we keep the transfer in a narrower
strip instead of projecting over the full scattering angle:
\begin{equation}
 \begin{gathered}
 \mathcal S_{ABjn}=\frac1{32\pi}\int_A^B ds\,w^{\rm tap}_{ABn}(s)
       \int_{-T}^0dt\,h_j(t)\mathscr E(s,t)=0,\\
 w^{\rm tap}_{ABn}(s)=\frac{(1-x_s^2)P_n(x_s)}{B-A},\qquad
 x_s=\frac{2(s-A)}{B-A}-1,\qquad
 h_j(t)=\frac1T P_j(1+2t/T),\qquad T=\frac43.
 \label{app:eq:certified-strips}
\end{gathered}
\end{equation}
The two windows are $[A,B]=[12,24]$ and $[24,48]$; the polynomial
orders are listed in Table~\ref{app:tab:certified-banks}. The taper
$1-x_s^2$ vanishes at the window endpoints. The subtraction term is
$-\delta_{j0}(2\delta_{n0}/3-2\delta_{n2}/15)g_0^{\rm m}$ and is
eliminated using the same coupling map. No positive-transfer strip
is used. The symmetric-point kernels at $t=4/3$ enter the coupling
maps and crossing equations instead.

The strip kernels also contain real-part contributions at arbitrarily
high even spin; these terms are included in the remainder bounds below.
The principal value at $\mu=s$ is evaluated analytically in these
smeared kernels. Apparent poles from crossed subtraction terms are
removed using the corresponding Legendre divided differences. This
is relevant both to accurate evaluation and to rigorous integration;
discarding an endpoint or a pole from a numerical grid would not
implement the stated equation. The full kernel formulas, including
these continuous limits, accompany the exact multiplier data.

\subsubsection{Certifying a supporting inequality}
\label{app:massive-certification}
For a direction $(n_0,n_2)$, the search selects real multipliers
$\lambda_i$ for the retained null equations
$N_i=\langle u_ir+v_ib\rangle_{\rm m}=0$. With
\begin{equation}
 U=n_0 u_0+\sum_i\lambda_i u_i,\qquad
 V=n_0 v_0+n_2 p_\ell+\sum_i\lambda_i v_i,
 \qquad
 n_0 g_0^{\rm m}+n_2 g_2^{\rm m}
 \le\left\langle\sqrt{U^2+V^2}+V\right\rangle_{\rm m},
 \label{app:eq:certified-support}
\end{equation}
unitarity has been maximized exactly at each energy and spin. This
is the support of the Minkowski sum used in the main text. The
optimization finds a useful inequality; certification checks the
chosen inequality without rerunning that optimization.

Each saved floating-point coefficient is interpreted as its exact
binary rational value. Rational combinations of witnesses are kept
rational. The target normal, all null multipliers and the final
upper constant are therefore unambiguous. The plotted region uses
an outward-rounded constant $C_{\rm plot}\ge C_{\rm cert}$ for each
inequality $n_0 g_0^{\rm m}+n_2 g_2^{\rm m}\le C_{\rm cert}$.

The integrals are evaluated with Arb interval arithmetic, which
bounds rounding errors as well as truncation errors. For the spins
integrated directly, $\beta=\sqrt{1-4/\mu}$ maps the whole spectral
cut to $0\le\beta\le1$. Exact rational partitions cover this closed
interval. Ordinary cells are bounded by interval or Taylor
enclosures; threshold caps, logarithmic source endpoints and the
infinite-energy endpoint receive separate analytic bounds.
Moments of the smearing weights and the finite series used to
evaluate the kernels have explicit remainder estimates. Thus no quadrature remainder is inferred from
agreement between two numerical grids.

The spin sum is divided as follows:
\begin{equation}
 \begin{array}{c|c|c|c}
 &\text{Direct integration}&\text{Finite remainder}&\text{Infinite remainder}\\\hline
 \text{Set I}&\ell=0,2,4,6&\ell=8,10,\ldots,40&\ell\ge42\text{ even}\\
 \text{Set II}&\ell=0,2,\ldots,40&\ell=42,44,\ldots,448&\ell\ge450\text{ even}
 \end{array}
\end{equation}
Every even spin is included; the split specifies how its contribution
is bounded. The
remaining integrals are controlled by a simple consequence of the
same disk support:
\begin{equation}
 V\le-\kappa p_\ell,\quad\kappa>0
 \quad\Longrightarrow\quad
 0\le\sqrt{U^2+V^2}+V\le\frac{U^2}{2\kappa p_\ell}.
 \label{app:eq:certified-tail}
\end{equation}
The proof bounds $V/p_\ell$ away from zero and integrates a bound on
$U^2/p_\ell$, including its infinite spin sum. In set II, compact
energies $4<\mu\le49$ are covered by finite-spin checks and analytic
large-spin estimates. Above all external-energy windows, at $\mu\ge49$, the variables
$\tau=\ell(\ell+1)/(\mu-4)$ and $\eta=1/[\ell(\ell+1)]$
allow interval bounds uniform in spin. The infinite-spin panels
include $\eta=0$, and a separate analytic estimate covers
$\tau\ge256$. The finite bridges between these domains are included
in the saved proof records. This is how the certificate covers
arbitrarily high energy and spin.

Production and higher-precision replays use the saved partitions;
the selected upper bounds are retained conservatively. The final
constant is an exact rational sum of the upper bounds for the direct
integrals, endpoint terms, finite-spin remainders and infinite-spin
remainders. For a positive
combination of witnesses, convexity of the disk support bounds the
untouched contributions by the corresponding combination of parent
costs. Selected low spins can then be integrated again to sharpen
that bound. Only the authenticated parent contributions for those
same spins are replaced, so no domain is omitted or counted twice.

\paragraph{Selection against the published dual bounds.}
For a direction $n=(n_0,n_2)$, an inequality improves the comparison
when its outward-rounded constant is smaller than the published
support in that same direction. For 24 inequalities we use
\begin{equation}
 C_{\rm plot}<\max_{g\in D_{\rm ref}}(n_0g_0^{\rm m}+n_2g_2^{\rm m}),
 \label{app:eq:improving-selection}
\end{equation}
where $D_{\rm ref}$ consists of the observed digitized dual branches
of Ref.~\cite{r26}. This comparison uses no approximate continuation
of their tips. The remaining inequality is
$g_0^{\rm m}\le2.705955$, compared separately with the published
extrapolated scalar bound $g_0^{\rm m}\le2.7272$ of Ref.~\cite{r5}.
The comparison with a figure has its graphical accuracy; the
certification of our inequalities does not depend on that figure.
Each retained inequality improves at least one direction. Their
intersection need not improve the entire published curve uniformly.

\paragraph{Results and verification files.}
These 25 numerical inequalities, together with positivity, give
\begin{equation}
 -7.751502\le g_0^{\rm m}\le2.705955,\qquad
 0\le g_2^{\rm m}\le1.052351.
 \label{app:eq:massive-resummed-endpoints}
\end{equation}
The full enclosure is the intersection of these half-planes and is
an outer bound on the amplitude region. The published dual and
primal curves use the same normalization,
$(c_{0,0},c_{1,0})=(g_0^{\rm m},g_2^{\rm m})$. The lower-panel
markers remain unconnected; nearby markers may be omitted for
legibility without removing their inequalities from the analysis.

The ancillary files contain the exact normals and constants, all
5,010 effective null multipliers, and a comparison table for these
25 inequalities. A minimal verification implementation accompanies
the kernel code, source moments, interval partitions, remainder
bounds and certificate assemblies. The public checking and search
commands accept only the retained inequalities. Parent witnesses
and shared intermediate records are included where their costs are
needed in a retained certificate; they are not additional reported
bounds.

Verification has three stages: check the exact data and their source
bindings; reassemble each inequality from its saved proof components;
then rebuild source moments and replay interval calculations. Hash
checks alone establish file integrity, not the inequalities. The
replay guide gives the commands and software versions, and the test
report distinguishes saved-component assembly from fresh interval
calculations. A separate search package contains both constraint sets
and the exact retained seeds, allowing the searches to be continued
without the proof archive. New search results require new
certificates.

\subsection{The forward Gram family and its limit}
\label{app:massive-forward-matrices}
In the forward limit we use the variable $x=2/(s-2)$, which runs from
$0$ to $1$ along the cut. The spectral average of
Eq.~\eqref{eft4d:eq:main-massive-measure} with the weight $F$ becomes
an integral with the measure
\begin{equation}
 d\mu^{\rm m}=\frac4\pi\frac{x(1+x)}{\sqrt{1-x^2}}\,dx,
 \qquad
 \mu_k^{\rm m}=\frac2\pi\left[
 B\!\left(\frac{k+2}2,\frac12\right)
 +B\!\left(\frac{k+3}2,\frac12\right)\right],
 \label{app:eq:massive-reference}
\end{equation}
where $\mu_k^{\rm m}$ are the moments of this measure and $B(a,b)$ is
the Euler beta function. With this reference measure the block Hankel
construction of Sec.~\ref{app:hankel-gram} applies without change. A
smaller matrix is obtained with weights adapted to the harmonics. We
define the two-component vectors
$e_j=(\cos j\vartheta_{\rm m},\sin j\vartheta_{\rm m})^T$ and choose
the weights
\begin{equation}
 \psi_0=\sqrt{p_\ell}\binom10,\qquad
 \psi_j=\frac F{\sqrt{p_\ell}}v_j,\qquad
 v_j=e_1+(-1)^je_j,\quad j=2,\ldots,N.
 \label{app:eq:massive-phase-weights}
\end{equation}
The entries of the Gram matrix
$\langle\psi_0^TW_\ell\psi_j\rangle_{\rm m}$ that mix $\psi_0$ with
$\psi_j$ are $\widetilde H_1(0)+(-1)^j\widetilde H_j(0)$. They are
equal to $\widetilde H_1(0)$, because the harmonics with $j\ge2$
vanish. The first diagonal entry is
$\langle p_\ell b_\ell\rangle_{\rm m}=g_2^{\rm m}$. Since
$W_\ell\preceq2I$, the remaining block is bounded from above by
$2\langle(F^2/p_\ell)[v_i\cdot v_j]\rangle_{\rm m}$. This reference
matrix contains a sum over all even spins, which we bound with the
inequality
\begin{equation}
 \sum_{\ell\ {\rm even}}\frac{2\ell+1}{P_\ell(\bar s/(s-4))}
 \le1+\frac{5(s-4)^2}{\bar s^2}+\frac{3(s-4)^2}{32}.
 \label{app:eq:reciprocal-legendre}
\end{equation}
The first two terms on the right-hand side bound the contributions of
the spins $\ell=0$ and $\ell=2$. For $\ell\ge4$ we use that the
Legendre polynomial is bounded from below by the quadratic term of
its Taylor expansion around the point where its argument is equal to
one, together with the sum
$\sum_{\ell\ge4,\ {\rm even}}(2\ell+1)/[\ell(\ell+1)(\ell-1)(\ell+2)]=1/24$.
The result is
\begin{equation}
 \begin{pmatrix}g_2^{\rm m}&\widetilde H_1(0)\mathbf1^T\\
 \widetilde H_1(0)\mathbf1&2T_N\end{pmatrix}\succeq0,
 \qquad \widetilde H_1(0)^2\le2K_Ng_2^{\rm m},\qquad
 K_N=(\mathbf1^TT_N^{-1}\mathbf1)^{-1},
 \label{app:eq:massive-gram}
\end{equation}
where $\mathbf1$ is the vector with $N-1$ components equal to one.
The matrix $T_N$ has the indices $i,j=2,\ldots,N$ and the entries
\begin{equation}
 \begin{gathered}
 (T_N)_{ij}=\frac1\pi\int_0^1
 \frac{w(x)D_{ij}(2x^2-1)}{\sqrt{1-x^2}}\,dx,\\
 w(x)=\frac{16(x+1)(x+3)}{27x}
       (371x^4-660x^3+426x^2-36x+27),\\
 D_{ij}(c)=1+(-1)^jT_{j-1}(c)+(-1)^iT_{i-1}(c)
                   +(-1)^{i+j}T_{|i-j|}(c).
 \end{gathered}
 \label{app:eq:massive-gram-reference}
\end{equation}
Here $T_j(c)$ denotes the Chebyshev polynomial of the first kind with
the argument $c$.
Each function $D_{ij}(2x^2-1)$ contains a factor $x^2$, which cancels
the factor $1/x$ in $w(x)$. 

The correction term in Eq.~\eqref{app:eq:massive-forward-map}
satisfies $-128p_\ell/27\le C_\ell\le-64p_\ell/27$. To prove this we
note that the ratio $C_\ell/p_\ell$ is an affine function of the
inverse of the Legendre polynomial $P_\ell(\bar s/(s-4))$, and this
inverse lies between $0$ and $1$. It is therefore sufficient to check
the inequality at these two values. Since $b_\ell\ge0$, the relation
$g_0^{\rm m}=\widetilde H_1(0)+\langle C_\ell b_\ell\rangle_{\rm m}$
then bounds the difference $g_0^{\rm m}-\widetilde H_1(0)$ between
$-128g_2^{\rm m}/27$ and $-64g_2^{\rm m}/27$. Together with
Eq.~\eqref{app:eq:massive-gram} this proves
\begin{equation}
 -\frac{128}{27}g_2^{\rm m}-\sqrt{2K_Ng_2^{\rm m}}
 \le g_0^{\rm m}\le
 -\frac{64}{27}g_2^{\rm m}+\sqrt{2K_Ng_2^{\rm m}}.
 \label{app:eq:massive-gram-bound}
\end{equation}
The first constants of the sequence are
\begin{equation}
 K_2=\frac{4496}{27}+\frac{5184}{5\pi}=496.5422085\ldots,
 \qquad K_4=286.0222386\ldots,\qquad K_{32}=235.8219144\ldots .
 \label{app:eq:massive-gram-sequence}
\end{equation}
They follow from the inverse matrix in
Eq.~\eqref{app:eq:massive-gram}, whose entries are known in closed
form.

The limit $N\to\infty$ can also be computed exactly, in analogy with the massless case. We identify the
vector $e_j$ with the complex number $e^{ij\vartheta_{\rm m}}$, and we
define the point $\zeta=-e^{i\vartheta_{\rm m}}$ of the unit circle,
which is not the function of Eq.~\eqref{app:eq:disk-variable}. A
linear combination $\sum_{j=2}^Na_jv_j$ with real coefficients $a_j$
and $\sum_ja_j=1$ is then equal to
$e^{i\vartheta_{\rm m}}(1-\zeta)Q(\zeta)$, where $Q$ is a polynomial
of degree at most $N-2$ with $Q(0)=1$. The constant $K_N$ is the
minimum of the quadratic form of the matrix $T_N$ on the vectors with
$\sum_ja_j=1$. It is therefore the smallest weighted norm of such
polynomials on the unit circle. The factorization of the positive
weight gives
\begin{equation}
 K_\infty=\exp\!\left[\frac1\pi\int_0^\pi\log V(\eta)\,d\eta\right]
 =233.7516883972\ldots,
 \qquad V(\eta)=2x^2w(x),\quad x=\sin(\eta/2),
 \label{app:eq:massive-gram-limit}
\end{equation}
where $\eta=\pi-\vartheta_{\rm m}$ is an integration variable. The
polynomial factor of $w(x)$ has the positive coefficients
$(27,18,80,48,128)$ in the Bernstein basis on the interval
$0\le x\le1$. The function $V$ is therefore positive except for a
zero at the endpoint $\eta=0$, and $\log V$ is integrable. Jensen's
inequality shows that the right-hand side of
Eq.~\eqref{app:eq:massive-gram-limit} is a lower bound for the
weighted norm, and polynomial approximations of the inverse of the
outer function reach this bound. This also proves that $K_N$
decreases monotonically to $K_\infty$. The limiting bound is an
analytic outer bound. It is not the exact region, because the bounds
$W_\ell\preceq2I$ and Eq.~\eqref{app:eq:reciprocal-legendre} were
used. In particular, it does not bound $g_2^{\rm m}$ from above. In
this family the number of nulls and the size of the matrix grow
together with $N$.

\end{document}